\documentclass[reprint, amsmath, amssymb, aps]{revtex4-2}

\usepackage{graphicx}
\usepackage{dcolumn}
\usepackage{bm}
\usepackage[colorlinks=true,citecolor=blue,linkcolor=blue,urlcolor=blue]
{hyperref}
\usepackage{amsmath}

\begin{document}  
\title{\Large Boltzmann Meets Lorentz: A Surrogate Model for Black Hole Echoes}

\author{Randy S. Conklin}
\email{randysconklin@gmail.com}
\affiliation{Department of Physics, University of Toronto, Toronto, Ontario, Canada,  M5S 1A7}

\author{Niayesh Afshordi}
\email{nafshordi@pitp.ca}
\affiliation{Department of Physics and Astronomy, University of Waterloo,
200 University Ave W, N2L 3G1, Waterloo, Canada}
\affiliation{Waterloo Centre for Astrophysics, University of Waterloo, Waterloo, ON, N2L 3G1, Canada}
\affiliation{Perimeter Institute for Theoretical Physics, 31 Caroline Street North, Waterloo, Ontario, N2L 2Y5, Canada}

\begin{abstract}
The existence of black hole horizons has not been strictly proven observationally, and indeed it may not be possible to do so. However, alternatives may be established by the observation of gravitational wave echoes that probe possible near-horizon structure. These echoes are proposed to be generated in exotic compact objects that are horizonless and feature a partially reflecting ``wall'' inside their light rings, creating a cavity in which gravitational perturbations may echo, while leaking out through the angular momentum barrier with each pass. The characteristic signature of echoes is a comb of nearly evenly spaced spectral resonances. While approximately true, deviations from this simple picture can lead to severe observational signal losses. In this paper, we explore such subtleties with the latest results for echo sourcing and geometry. A physically motivated echo model is then developed as a sum over Lorentzian spectral lines, parametrized by functions of the horizon frame frequency and the size of the cavity. Our final spectrum is a function of only the mass and spin of the black hole, as well as the UV scale of the near-horizon physics. We then apply this model in a search for echoes in the gravitational wave event with the loudest ringdown signal in LIGO/Virgo, i.e. GW190521. We interpret our findings as a measurement of the fractional energy in post-merger echoes equal to $E_{Echoes} / E_{GR} = 8.9 \pm 4.5\%$, where the uncertainty range represents the 90\% credible region. The robustness of this result is tested against noise backgrounds and simulated injections, and we find that a signal persists through modifications to the model and changes in the data search.
\end{abstract}
\maketitle

\section{Introduction}

A complete theory of quantum gravity has not been established, partly due to Planck scale experimentation seemingly being out of reach. It is nevertheless expected that black hole horizons may be replaced by quantum effects such that they are no longer perfectly absorbing, and that this may lead to a macroscopic signal in the form of gravitational wave echoes, bringing the Planck scale into the reach of observations \cite{S_Cardoso_89, S_Abedi_18}.

There are several motivations for this suggestion, one of significance being the information paradox. If black holes are actually horizonless objects, then unitarity can be preserved and the issue evaporates. There is also the proposal of objects that obey general relativity as normal but in a special configuration with exotic matter (the gravastar) \cite{S_Mazur_a}. Whatever the case, the test for echoes is a test for new physics, and with their distinct spectral signature, echoes present an interesting search target for gravitational wave observatories.

Gravitational wave echoes occur, in the simplest case, whenever there is a gravitational cavity containing an oscillating perturbation. For the signal to escape and reach an observer, one side of the cavity should be leaky such that upon contact with it part of the perturbation transmits and travels towards infinity. Therefore, in an augmented black hole geometry, with a classical angular momentum barrier surrounding a (non-classical) reflecting ``wall'' echoes may be produced by perturbing the background. 

The wall is different for each theory of quantum gravity, and even for different exotic compact objects (ECO's) within each theory. For example, the wall interior to the angular momentum barrier for a wormhole is the angular momentum barrier on the other side of the origin. Of particular interest for the present paper is the recent proposal that generic quantum considerations may lead to a Boltzmann reflection of frequencies at the wall, with temperature dependent on the spin of the ECO \cite{S_Wang_40}. Physical perturbative models require absorption at the wall to counteract the runaway growth of superradiance for spinning backgrounds, which occurs for frequencies: $\omega < m \Omega_H$, where $\Omega_H$ is the superradiance frequency defined by the mass $M$ and spin $a$ of the background, and $m$ is the azimuthal harmonic number \cite{S_Brito_b}. Perturbations must therefore dissipate at each contact with the wall.

The angular momentum barrier acts as a high-pass filter, and perturbations do leak out upon contact with it. If the source of the perturbation is a single local outgoing wavepacket inside the cavity, a high-frequency pulse characterized by the fundamental quasinormal mode (QNM) frequency \cite{S_Berti_j} of the background, and higher frequencies that are present, will be emitted. After this, each oscillation of the perturbation will occur over one time delay $t_d \approx 2|x_0|$, where $|x_0|$ is the distance in the tortoise coordinate between the wall and the angular momentum barrier.

The formalism developed so far for echoes resides in the linear regime where perturbation theory is valid. For compact merger events, as observed by LIGO and Virgo, the inspiral and merger phase provide the source of perturbations that lead to ringdown and possibly echoes. Given that this sourcing is outside of the perturbative regime, a relation must be established connecting this to echoes. Recently, advances were made here as sources were developed for infalling particles on spinning backgrounds (compared to static backgrounds \cite{S_Mark_79}) and a link between this and echoes was established \cite{S_Micchi_91, S_Xin_c}. This represented significant progress as it was also recently discovered that source functions must be carefully defined to quantify echo distortion, and previous models were often simpler than required in the physical picture \cite{S_Conklin_24, S_Conklin_38}. In this paper, we take the further step of sourcing echoes with surrogate models for gravitational wave signals from realistic binary mergers \cite{S_PyCBC_d}. Indeed, in our calculation for the energy potentially emitted through echoes in GW190521, we measure the GR event energy and compare it to the measured energy of the echoes partly by sourcing the echoes with the waveform that reproduces the main event.

In Fourier space, echo resonances present their most striking feature  - a series of potentially dozens of sharp poles that are approximately evenly separated by $1 / t_d$. This spectrum resembles a comb with teeth of various lengths, a unique pattern that may be used as a target signal in data searches. 

This has been done with proposed success in \cite{S_Conklin_73}, where a series of trapezoids of varying width but otherwise equal shape were used to search the LIGO O1 data, and upper limits on p-values for detection were given. The most recent work on this updates the original trapezoids to become an evenly spaced series of equal triangles \cite{S_Holdom_90}. In this model, the optimal bandpass and number of echoes were inferred \textit{a posteriori} from the data, and a source was chosen as a Gaussian distribution of frequencies about the horizon frame frequency in the proposed context of energy minimization. The background for this model was a truncated Kerr black hole with near perfect energy reflection at the wall, and $\varepsilon$ damping to counter instability.

In this paper, we build upon previous work by generating physically motivated echo spectra for a variety of geometric parameters by incorporating the Boltzmann boundary condition and realistic sources, and by representing echo resonances as a sum of Lorentzians according to the shapes of their quasinormal modes (QNM's), with each resonance defined by its phase, amplitude, width, and location. We then proceed to study these features in detail. 

For example, we quantify the departure of the separation between resonances away from constant integer multiples of $\sim 1/t_d$ (as has often been assumed in the literature), and write a more accurate description as a Taylor expansion in the horizon frequency $k= \omega -m\Omega_H$. This nonconstant separation was first noticed as a 1-2\% departure from constancy peaking near the horizon frequency in \cite{S_Conklin_73}, and part of our work here has been to model this efficiently and with significant precision for physically motivated echoes. Because the locations of these resonances are critical for optimal echo detection, we provide a functional form for them valid for all relevant binary merger parameters. 

After developing our model, we advocate an optimal search strategy that involves comparing the results of the signal search to noise backgrounds and injections while accounting for LIGO/Virgo/KAGRA measurement uncertainties through Monte Carlo parameter chains, avoiding {\it a posteriori} statistics. Finally, as an example, we use this method to measure the energy in post-merger echoes in GW190521, which is the LIGO/Virgo event with loudest ringdown signal \cite{S_Abbott_e}. 

This paper is divided into three main sections. In Section \ref{sec:Setup}, we introduce the theoretical background and terms that will be useful throughout the paper. We then continue into Section \ref{sec:TheModel} where we construct the model, calibrate its parameters, and produce a surrogate model optimized for numerical efficiency and accuracy. To apply this in a data search, we continue into Section \ref{sec:GW190521} where we examine GW190521, a promising candidate for echoes given its strong ringdown. By the end of the paper, the reader will have been provided with a calibrated surrogate model with parameters explicitly given, as well as the first sample of a data search using this tool along with a statistical analysis of the results. These points are summarized in the conclusion.

\section{Setup}\label{sec:Setup}

Echoes are clearly distinguished by their representation in frequency space, and therefore we begin by defining their spectrum $\psi_\omega$ that would be measured by an asymptotic observer as a function of frequency $\omega$.  In our formalism, where $x$ is a tortoise coordinate, the location of the observer is approaching $x \to \infty$.

For easy access to the most relevant equations, we start from a high-level perspective showing the primary functions of significance, and then proceed to explain each term. The following subsections explain the theory in more detail.

Echoes are gravitational wave perturbations, in this context coming from disturbed exotic compact objects (ECOs). Their dynamics are governed by the Teukolsky differential equation \cite{S_Teukolsky_f}, and the frequency spectrum that is a solution to this may be written as (in a transformed notation called the Sasaki-Nakamura (SN) formalism \cite{S_Sasaki_g})
\begin{equation}
\psi_\omega = e^{i\omega x}KS ,
\end{equation}
where $K$ is called the transfer function. In a way, $K$ is the most important functional form that describes echoes. This is because echoes can almost entirely be described by a sum of sharp frequency resonances, and these are encoded in $K$. Here, $S$ is the source function generated by initial conditions and/or active sourcing, and this plays the important role of modifying the amplitudes of the sum of resonances \cite{S_Conklin_24, S_Conklin_38}. 

The transfer function has multiple representations in the literature. We choose to write \cite{S_Conklin_73}
\begin{equation}
K = \frac{T_{\textrm{BH}}(\omega)}{1 - R_{\textrm{BH}}(\omega) R(\omega) e^{-2i k x_0}} ,
\end{equation}
where $k = \omega - m\Omega_H$ is the horizon frequency and $x_0 = t_d/2$ is the cavity size (the tortoise coordinate distance between the wall and the angular momentum barrier). $T_{\textrm{BH}}$ and $R_{\textrm{BH}}$ are standard general relativistic quantities that we call the transmission and reflection coefficients. For perturbations with angular momentum, which are those of interest here, they give, respectively, the square root of the ratio of transmitted and reflected energies across the angular momentum barrier of the black hole or ECO. They are independent of the boundary condition at the wall and the size of the cavity. $R(\omega)$ is the square root of the ratio of reflected to incoming energy at the wall, sometimes called $R_{\textrm{wall}}$ in the literature.

Boltzmann reflection, which will be of particular interest in this paper, is defined by $R = e^{-\frac{|k|}{2\alpha T_\textrm{H}}}$, where $T_\textrm{H} = \frac{\sqrt{1 - a^2}}{4\pi M(1 + \sqrt{1 - a^2})}$ is the ECO temperature and $\alpha$ is a number \cite{S_Wang_40}. The standard argument sets $\alpha = 1$ for generic quantum modifications near the horizon. However, up to $\alpha = 2$ is energetically permitted. In this paper, we focus on the standard case, though we do also search the data with the extremal value since it results in a much wider range of significant resonances given the reduced suppression that leads to a potentially stronger signal. The superradiance frequency $m\Omega_H$ defines the center of the peak signal region for Boltzmann echoes, given that at this frequency there is no absorption at the wall. Whether an actual resonance appears at this frequency or whether the highest resonances are immediately adjacent to this depends on the net combination of phase and amplitude factors, as we will discuss.

The motivation for this definition of the transfer function is largely that it has a sensible interpretation for Boltzmann echoes in the geometric optics approximation, as is described in the following paragraph. It is also simply derived as the inverse of the position-independent Wronskian for the gravitational perturbation equation, and so has a concise mathematical definition \cite{S_Conklin_73}.

The transfer function has a very useful interpretation in which the denominator is replaced by a geometric expansion in $P \equiv R_{\textrm{BH}}Re^{-2i k x_0}$, such that \cite{S_Mark_79}
\begin{equation}
K \approx T_{\textrm{BH}}(1 + P + P^2 + ...) ,
\end{equation}
valid when $|P| < 1$. This representation is called the geometric optics approximation. For Boltzmann echoes, which are of primary interest in this paper, the inequality holds everywhere except at the superradiance frequency ($k = 0$) where $|P| = 1$. In contrast to Boltzmann echoes, other models often have $|P| \geq 1$ due to ergoregion instability (for example the perfect energy reflection and Dirichlet models) where this representation breaks down. 

Because of this, for Boltzmann echoes we may interpret the transfer function as an expansion in echo pulses where each term in the expansion in $P$ is an echo \cite{S_Mark_79}. In this picture, for an initial perturbation originating inside the cavity between the wall and the angular momentum barrier, the first pulse to leave the cavity (generating the ringdown) has frequency content $e^{i\omega x}T_{\textrm{BH}} S$, while the second has $e^{i\omega x}T_{\textrm{BH}} PS$ and so on. The interpretation of these terms as echoes is sensible since the first echo (the burst after the initial pulse) results from first reflection off the angular momentum barrier, then reflection off the wall, then transmission through the barrier, and subsequent echoes pick up additional reflection factors encoded by $P$. 

Later, we will find it useful to separate the GR event from the echo waveform, which is very easy with this interpretation. The echo component of the spectrum may be subtracted from the total waveform by simply removing the initial pulse, which is the ringdown, such that
\begin{equation}
\begin{split}
\psi_{\omega, echo} &= e^{i\omega x}K_{echo} S \\
&= e^{i\omega x}(K - T_{\textrm{BH}}) S \\
&= e^{i\omega x}\frac{T_{\textrm{BH}}(1 - (1 - P))}{1 - P}S \\
&= e^{i\omega x}\frac{T_{\textrm{BH}}P}{1 - P}S.
\end{split}
\end{equation}

As a final introductory word, we qualitatively describe the echo spectrum in terms of the transfer function. We mentioned earlier that echoes in Fourier space are sums of sharp resonances. These resonances typically occur when $K$ is maximized, which generally implies the denominator of $K$ being minimized. In the simplest picture, where $|P| \approx 1$ is always true, resonances are separated by roughly $1/t_d$, corresponding to the periodicity of the phase factor in the exponential. Hence, the transfer function embodies the characteristic structure of echoes which is a comb of resonances separated by a function of the inverse time delay. This all remains essentially true in the more realistic picture, where $P$ has more magnitude and phase structure than in this toy description, and the numerator can modify the amplitudes of resonances through the source function.

\subsection{Teukolsky and Strain}

Here we proceed to describe the theory in more detail, beginning with the equation from General Relativity that describes the motion of perturbations on a background gravitational geometry. The essential new feature in this formalism compared to that of standard GR is the introduction of a boundary condition (the exotic wall) interior to the angular momentum barrier and its consequences.

The Teukolsky equation for $s = -2$ governs the dynamics of gravitational perturbations on a Kerr background of mass $M$ and spin $a$. In general form in vacuum it is \cite{S_Teukolsky_f}
\begin{equation}\label{eq:Teuk}
\begin{split}
&\left[\frac{(r^2+a^2)^2}{\Delta}-a^2\sin^2\theta\right] \frac{\partial^2\psi}{\partial t^2} + \frac{4Mar}{\Delta}\frac{\partial^2\psi}{\partial t\partial\phi} \\
&+\left[\frac{a^2}{\Delta}-\frac{1}{\sin^2\theta}\right]\frac{\partial^2\psi}{\partial\phi^2} - \Delta^{-s}\frac{\partial}{\partial r} \left( \Delta^{s+1}\frac{\partial\psi}{\partial r}\right)\\
&-\frac{1}{\sin\theta}\frac{\partial}{\partial\theta}\left(\sin\theta\frac{\partial\psi}{\partial\theta}\right) - 2s\left[\frac{a(r-M)}{\Delta}+\frac{i\cos\theta}{\sin^2\theta}\right]\frac{\partial\psi}{\partial\phi} \\
&- 2s\left[\frac{M(r^2-a^2)}{\Delta} - r - ia\cos\theta\right]\frac{\partial\psi}{\partial t}\\
& + (s^2\cot^2\theta-s)\psi = 0 ,
\end{split}
\end{equation}
where $\Delta=(r^2-2Mr+a^2)$. To generate echoes from this equation, an exotic boundary condition and a source term must be added. For Boltzmann echoes, which are stable because they have sufficient absorption, the Teukolsky equation may be immediately Fourier transformed. Other common reflectivities, such as those for perfect energy and Dirichlet boundary conditions, require Laplace transformation when $a \neq 0$ in the linear perturbative regime because of superradiance, since the Fourier transform is only defined when
\begin{equation}
\int_{t_0}^\infty |\psi|^2 dt < \infty,
\end{equation}
a condition that does not hold in the presence of instability \cite{S_Berti_h}. Causality is enforced in the Laplace (or Fourier) transform by initiating a start time $t_0$ such that
\begin{equation}
\mathcal{L} \psi(t) \equiv \widehat{\psi}(\omega) = \int_{t_0}^{\infty}\psi(t)e^{i\omega t}dt .
\end{equation}
Having a start time leads to non-operator initial condition terms in the differential equation. These may be lumped together with any active source term (such as an infalling particle) to eventually form what we call here the source term $S$ \cite{S_Conklin_24}. The value of $\omega$ in the Laplace transform is implicitly offset from the real axis to ensure all poles are on the side of the contour causing decay. The inverse of this transformation reveals this offset in the integration bounds:
\begin{equation}\label{eq:InverseLaplace}
\psi = \frac{1}{2\pi}\int_{-\infty + ic}^{\infty + ic}  \widehat{\psi} e^{-i\omega t} d\omega .
\end{equation}
Angular variables may be separated out of the Teukolsky equation, provided that the source term is properly handled \cite{S_Conklin_24}, leaving the radial operator and harmonic-indexed variable
\begin{equation}
\mathcal{R}R_{lm} = \left[\Delta^2\frac{d}{d r} \left(\frac{1}{\Delta}\frac{d}{d r}\right) - V(r)\right]R_{lm} ,
\end{equation}
where
\begin{equation}
V(r) = V = -\frac{K_T^2 + 4i(r-M)K_T}{\Delta} + 8i\omega r + \lambda ,
\end{equation}
$\lambda$ is the eigenvalue of the spin-weighted spheroidal harmonic equation left over from the separation \cite{S_Toolkit}, and
\begin{equation}
K_T = (r^2+a^2)\omega -ma .
\end{equation}
The Teukolsky radial variable $R_{lm}$ is indexed by the angular modes, and in this paper we will focus on the dominant $l = m = 2$ mode \cite{S_Oshita_17, S_Oshita_15}.

In a schematic notation, where angular components and normalization factors are excluded, the Teukolsky variable for $s = -2$ may be written as
\begin{equation}
\psi_{4,\omega} \sim R_{lm\omega} .
\end{equation}
This is directly related to strain $h$ (in time) by
\begin{equation}
\psi_{4} = \dfrac{1}{2} (\ddot{h}_{+}-i \ddot{h}_{\times}) , 
\end{equation}
and for this we define
\begin{equation}
\label{eq:strain}
h(t) = \dfrac{1}{2}(h_{+}-i h_{\times}) .
\end{equation}

\subsection{Homogeneous Solutions and the Sasaki-Nakamura Formalism}

Because we only require the asymptotic signal that makes its way to observers towards infinity, several simplifications occur. To discern these, it is helpful to first look at the asymptotic homogenous (source-free) solutions to the Teukolsky equation, and to use the Green's function technique to connect these to the full asymptotic signal \cite{S_Conklin_73}.

Because of its nonlocal potential, the homogenous solutions to the Teukolsky equation feature radial dependence in their asymptotics. Including the exotic boundary term from reflection off the wall, they are equal to
\begin{eqnarray}\label{eq:TEamp}
R_{lm\omega}\to
\left\{\begin{array}{cc}
B_\textrm{trans}\Delta^2 e^{-i k x}+B_\textrm{ref}e^{i k x},& x\to-\infty\\
B_\textrm{in}\frac{1}{r}e^{-i \omega x}+B_\textrm{out}r^3 e^{i \omega x},& x\to\infty . \end{array}
\right.
\end{eqnarray}
The radial dependence of these homogenous solutions on both sides of the angular momentum barrier leads to numerical difficulties, and various transformations are present in the literature to simplify these asymptotics for numerical efficiency. In this paper, we choose to work in the Sasaki-Nakamura formalism, described in detail in \cite{S_Sasaki_k}, where the asymptotics are simple travelling waves defined by the horizon frame frequency $k$ towards the horizon and $\omega$ towards infinity:
\begin{eqnarray}\label{eq:SNEamp}
X_{lm\omega}\to
\left\{\begin{array}{cc}
A_\textrm{trans}e^{-i k x}+A_\textrm{ref}e^{i k x},& x\to-\infty\\
A_\textrm{in}e^{-i \omega x}+A_\textrm{out}e^{i \omega x},& x\to\infty.\end{array}
\right.
\end{eqnarray}
In particular, we solve for these homogenous amplitudes using the SN differential equation, the local version of the Teukolsky equation, and use the Green's function technique to connect these to the source and the final result seen by an observer towards infinity.

The transformation between the Teukolsky and SN formalisms is well known in the literature \cite{S_Sasaki_k}. For our purposes, since we are just interested in asymptotic spectral amplitudes (the quantities just described), we are able to simply connect these two equivalent formalisms using the known homogenous amplitude conversion factors \cite{S_Conklin_73}
\begin{eqnarray}
& B_\textrm{in}=-\frac{1}{4 \omega ^2}A_\textrm{in},\quad
B_\textrm{out}=-\frac{4 \omega ^2}{c_0}A_\textrm{out}, \nonumber\\
& B_\textrm{trans}=\frac{1}{d}A_\textrm{trans},\quad
B_\textrm{ref}=\frac{1}{g}A_\textrm{ref} ,
\end{eqnarray}
with coefficients defined by
\begin{align}
c_0&=\lambda(\lambda+2)-12a\omega(a\omega-m)-i 12\omega M,\nonumber\\
d&=-4(2M r_+)^{5/2}\left[(k^2-8\epsilon^2)+i 6 k \epsilon\right],\nonumber\\
g&=\frac{-b_0}{4k(2Mr_+)^{3/2}(k+i 2\epsilon)},\\
b_0&=\lambda ^2+2 \lambda-96 k^2 M^2+72 k M r_+ \omega -12 r_+^2 \omega ^2\nonumber\\&-i [16 k M \left(\lambda+3-3\frac{M}{r_+}\right)-12 M \omega -8 \lambda  r_+ \omega],\nonumber\\
\epsilon&=(r_+-M)/(4M r_+) ,
\end{align}
where $r_+ = M + \sqrt{M^2 - a^2}$ is the outer horizon radius.

Putting all of this together, we can write a few simple relations connecting our variables. A general Teukolsky amplitude $Z$ is connected to strain $h$ by
\begin{equation}
Z \sim \omega^2 h_\omega e^{-i\omega x} .
\end{equation}
The relevant Teukolsky amplitude for observation, meaning for outgoing waves towards infinity, is related to the relevant Sasaki-Nakamura amplitude $X$ by
\begin{equation}
Z = -\frac{4\omega^2}{c_0} X.
\end{equation}
Putting all this together, the strain $h_\omega$ is then extracted from the Sasaki-Nakamura amplitude by
\begin{equation}
h_\omega \sim -\frac{4}{c_0} X e^{i\omega x} .
\end{equation}

The transmission and reflection coefficients in the transfer function may be defined by the homogenous solution satisfying the boundary conditions for waves towards infinity
\begin{eqnarray}
\psi_\textrm{right}(x)&\to&
\left\{\begin{array}{ll}
D_\textrm{trans}e^{-i k x}+D_\textrm{ref}e^{i k x},& x\to x_0\\
e^{i \omega x},& x\to\infty\end{array}
\right. 
\end{eqnarray}
according to
\begin{eqnarray}
T_{\textrm{BH}} = \sqrt{\frac{\omega}{k}}\left|\frac{b_0}{c_0}\right|\frac{1}{D_{\textrm{ref}}} \\
R_{\textrm{BH}} = \left|\frac{b_0}{C}\right|\frac{D_{\textrm{trans}}}{D_{\textrm{ref}}} ,
\end{eqnarray}
where
\begin{align}
|C|^2 = & \lambda ^4+4 \lambda ^3+\lambda ^2 \left(-40 a^2 \omega ^2+40 a m \omega +4\right) \nonumber\\
& + 48 a \lambda  \omega  (a \omega +m)\nonumber\\
& + 144 \omega ^2 \left(a^4 \omega ^2-2 a^3 m \omega +a^2 m^2+M^2\right) .
\end{align}

Some of the key quantities defined here are shown in Fig.\ref{fig:ConversionFactors} in the appendix, where we plot the functions that appear when transforming between formalisms and converting between energies and amplitudes. Further details are provided in the appendix of \cite{S_Conklin_73}. We also plot in Fig.\ref{fig:RBHTBH} in the appendix the reflection and transmission coefficients.

In summary, echoes are described by gravitational wave perturbations on background geometries containing black holes with exotic boundary conditions. Such perturbations are described by the Teukolsky equation. However, the Teukolsky equation has a nonlocal potential that leads to numerical challenges. Therefore we transform variables to obtain the Sasaki-Nakamura equation which is equivalent but with a localized potential. In this paper, background calculations are done in the SN formalism, and these are connected back to the Teukolsky equation through asymptotic homogenous spectral amplitude conversion factors, and through these to the observable strain by a final simple conversion.

\section{The Model}\label{sec:TheModel}

In Fourier space, echoes are characterized by a sum of sharp resonances. These correspond to complex poles in frequency, called quasinormal modes (QNM's), with the real part giving the oscillation rate of the wave and the imaginary part giving the exponential decay rate \cite{S_Berti_j}. The shapes of QNM's in frequency space are Lorentzian, and therefore we model echoes as a sum over Lorentzians. The model is parametrized by the location $l_r$, amplitude $a_r$, width $w_r$, and phase $\phi_r$ of the resonances such that the full model is the sum
\begin{equation}
\psi_\omega = \sum_{r=1}^n \frac{a_r}{\omega-l_r + iw_r}e^{i\phi_r} ,
\end{equation}
where $n$ is the number of resonances. The parameters in this model depend on the mass $M$, wall location $x_0$, and spin $a$.

Frequencies for echoes scale linearly with $M$, and it is possible to plot results as a function of the dimensionless quantity $\omega M$. We will often do this to maintain results generalizable to different gravitational wave events. To convert back to SI units from these dimensionless frequencies, we multiply by $c = 1/(4.9268*10^{-6}[s]M)$ with $M$ in units of solar masses. Gravitational wave data is usually plotted as a function of frequency in Hertz rather than $\omega$, so an extra division by $2\pi$ may be required for comparison. 

Quantum deviations away from GR are expected to occur around the Planck scale, and for this $x_0 \approx -450$ \cite{S_Micchi_2}. This estimate is approximated with the time delay $t_d$ where \cite{S_Abedi_65}
\begin{equation}\label{eq:TimeDelay}
2x_0 \approx t_d \approx \frac{4GM}{c^3}\left(1 + \frac{1}{\sqrt{1-a^2}}\right)\times \ln{\left(\frac{M}{M_P}\right)}.
\end{equation}
A typical value for observed binary mergers is $a \approx 0.7$, with the relevant range extending from about $a = 0.6$ to $a = 0.8$. For definiteness, the approximately Planckian value for the wall location and the mid-range spin will often be used in this paper, with the fuller ranges being explored during model construction.

In the following subsections, we discuss the calculation of the model parameters and we evaluate the model across a wide range of background geometry values to test its accuracy and relevance for echo searches.

\subsection{Location $l_r$}

Given their sharpness, the location of the resonances $l_r$ is the most sensitive parameter in this model, and small inaccuracies can lead to a significant loss of signal. We find that the resonances are not quite evenly spaced, but we can express the departure from constancy as a Taylor expansion about the superradiance frequency where Boltzmann echoes have their sharpest and highest resonances. 

\begin{figure}
\centering
\includegraphics[width=0.47\textwidth]{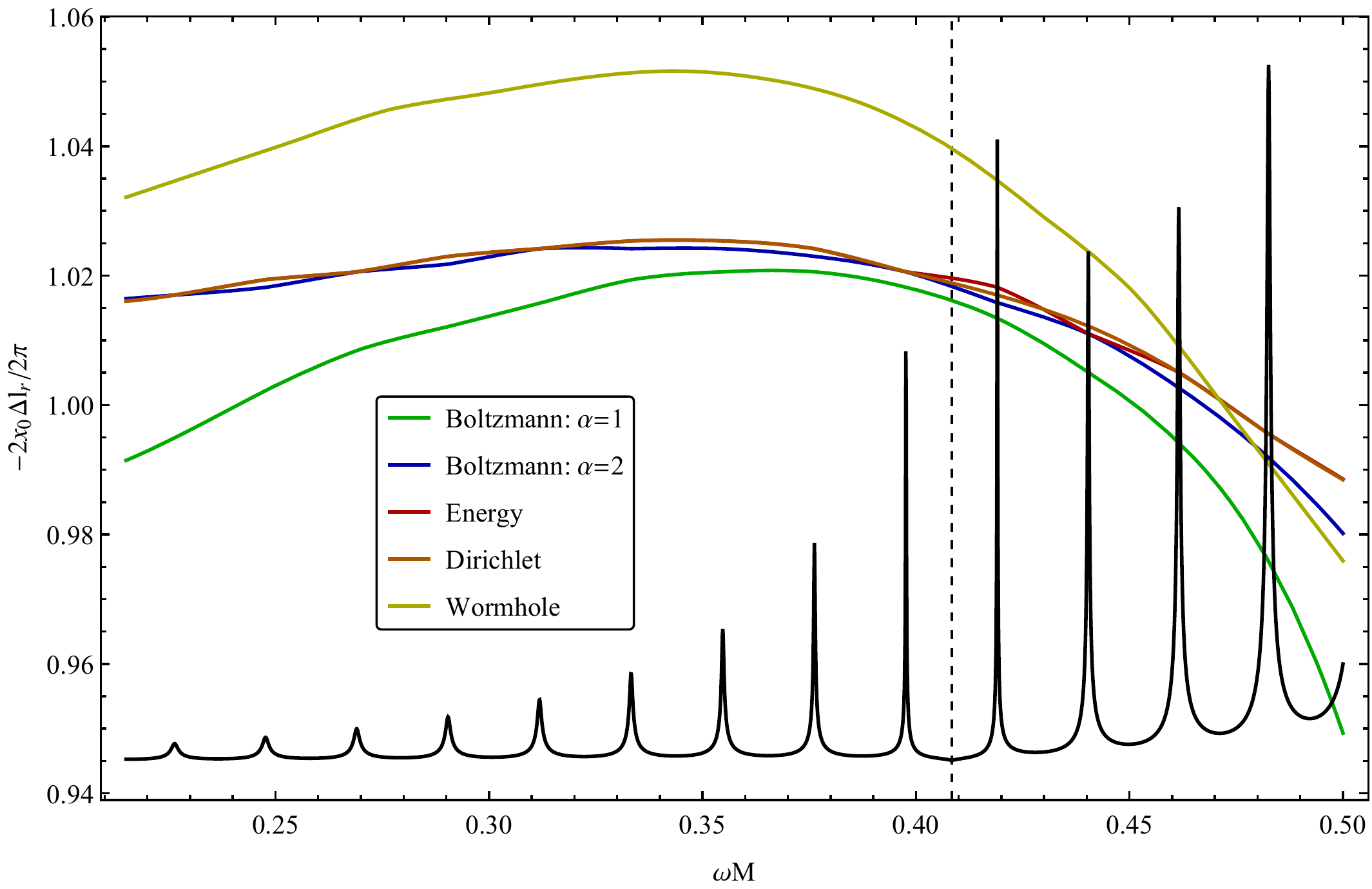}
\caption{\label{fig:EcoSeps} The five legended curves give the separation $\Delta l_r = l_{r+1} - l_r$ between resonances for each of the marked ECO's. The superradiance frequency $\omega_0 = m\Omega_h$ (or, equivalently, $k = \omega - \omega_0 = 0$) is marked by a vertical dashed black line. The resonances in the background (the black curve) are a sample spectrum, for perfect energy reflection, shifted vertically and resized for visualization. Here we set $a = 0.7$ and $x_0 = -150$. The $\alpha$ = 2 Boltzmann, Energy, and Dirichlet curves are barely distinguishable over much of this frequency range.}
\end{figure}

The separation of peaks $\Delta l_r = l_{r+1} - l_r$ is calculated by determining the location of the maximum of each resonance $l_r$. The sharper the resonances, the more consistently this may be determined as sharpness causes the peak locations to be increasingly independent of the source/initial conditions generating the echoes, as we will discuss. In Fig.\ref{fig:EcoSeps}, we plot the separation between resonances $l_{r+1} - l_r$ as a function of dimensionless frequency for echoes from several different ECO's.
The ``Energy'' curve gives the separation function for a wall boundary condition that perfectly reflects incoming energy. This is not a physical boundary condition when linear perturbation theory is assumed, since it leads to instability through superradiance, yet it is often studied for its simplicity, if one ignores excitation of the microstates of the black hole. The ``Dirichlet'' boundary condition perfectly reflects amplitudes rather than energy. This leads to additional conversion factors compared to the ``Energy'' reflection. These curves depart from constancy at up to several percent. For the ``Wormhole'', waves travel across the central point and reflect off the angular momentum barrier on the other side. The boundary condition for the wormhole is then equal to $R_\textrm{BH}$. Fig.\ref{fig:RBHTBH} gives the reflection and transmission rates for perturbations across the angular momentum barrier.

In this paper, we focus on the Boltzmann boundary condition, it arguably being the most physical, and here two points should be noted. First, the Boltzmann spectrum is localized to positive frequencies around the superradiance frequency, while the other boundary conditions in Fig.\ref{fig:EcoSeps} feature resonances extending to about $\omega M = -0.5$ \cite{S_Conklin_38}. This is due to the boundary condition which sharply peaks at the superradiance frequency with exponential suppression elsewhere. We focus on the standard $\alpha = 1$ Boltzmann model but also plot the $\alpha = 2$ variation which allows for more signal to be observed because of less suppression at the boundary. Second, with the exception of the wormhole, the Boltzmann boundary condition, especially for $\alpha = 2$, is comparable to that of the other boundary conditions over frequencies near $\omega_0$. In this way, up to quantifiable deviations, the separations of resonances derived in this paper are relevant over positive frequencies near the superradiance frequency for a variety of ECO's, and the model presented here may be used in data searches for alternative ECO models with a slight expansion of the priors.

\begin{figure}
\centering
\includegraphics[width=0.47\textwidth]{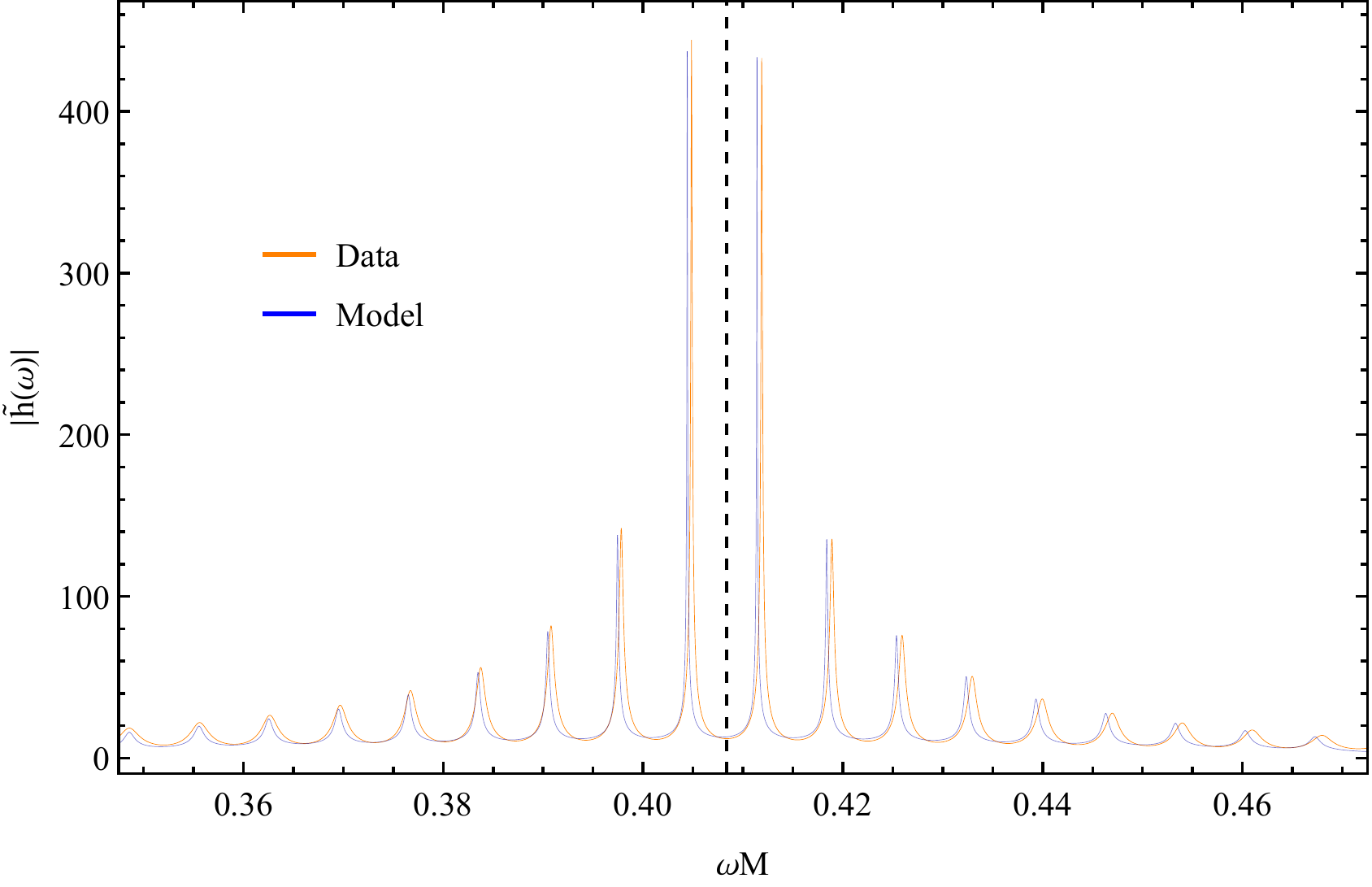}
\caption{\label{fig:SepDiff} Orange: The full numerical solution to the final waveform with $a = 0.7$ and $x_0 = -450$ using the GR surrogate source. These parameters and the source are consistent with the data search and will be discussed further throughout this paper. Blue: The model for the same parameters except assuming a constant separation between resonances, with the first resonance in this plot properly aligned. Both curves approximately cover the range of frequencies used in the data search. This demonstrates the nearly worst-case scenario where the incorrect separation function leads to only 38\% overlap compared to 97\% for the corrected model. On the other hand, with constant separation and the highest peak aligned (not shown here), the overlap goes up to 93\%. These results are for the standard Boltzmann model. For echoes with sharper resonances, such as for when $\alpha = 2$, the need for a corrected separation function becomes even more apparent. For $\alpha = 2$, the full model presented here gives 99\% overlap with the full solution. For the constant time delay assumption when the first resonance is aligned, the overlap drops to 24\%. For the best-case scenario under this assumption, when the highest peak is aligned, the overlap is 86\%. These numbers assume a time delay measured by the size of the cavity. For a data search with a flexible separation, these numbers could be made higher by aligning the top two peaks, with the trade-off being loss in accuracy of the time delay measurement.}
\end{figure}

In the final analysis, the deviation of the separation function away from constancy is important since sharp resonance peaks can be missed by not including this information. In Fig.\ref{fig:SepDiff} we plot the scenario where the lowest frequency peak is aligned and a constant separation between peaks is assumed.
This leads to a misalignment of the primary resonances causing the overlap integral of the plotted full solution with the constant-separation surrogate model to be 38\%. This is to be compared with the value of 97\% for the corrected model used in this paper. Though this number drops further as the properly aligned peak goes to lower frequencies away from the peak region, and especially if all of the peaks are misaligned, we provide these percentages as conservative estimates relevant for data searches where the frequency band may be relatively narrow compared to the full spectrum. The best case scenario, while still assuming a constant separation, would be attained by aligning the primary peaks and letting the less important auxiliary peaks to be misaligned. For standard $\alpha = 1$ Boltzmann echoes where the two peaks adjacent to the superradiance frequency dominate, little signal is lost with the overlap integral being as high as 93\%. The reader considering these numbers must determine where in the 38\% - 93\% range their overlap might land, and whether it may be better to include nonlinearities such as in the separation function presented in this paper. Whatever the case, the best solution is to account for the nonconstancy, even if this is only an improvement in signal inclusion by several percent. It should be noted, however, that the upper end of this range is a limiting case that is unlikely to be found in many other searches. If, for example, there are more than a few resonances that dominate, this value will decrease dramatically. Of particular note are models that feature resonances at positive and negative frequencies, where misalignment over the large range of frequencies may be fatal. Additionally, for models with wall absorption significantly less than that of the Boltzmann model, the auxiliary peaks can take on more importance causing the overlap to drop. For example, for the $\alpha = 2$ Boltzmann model, the same overlap range drops to 24\% - 86\%, compared to the overlap of the corrected model presented here which is 99\%.

\begin{figure}
\centering
\includegraphics[width=0.47\textwidth]{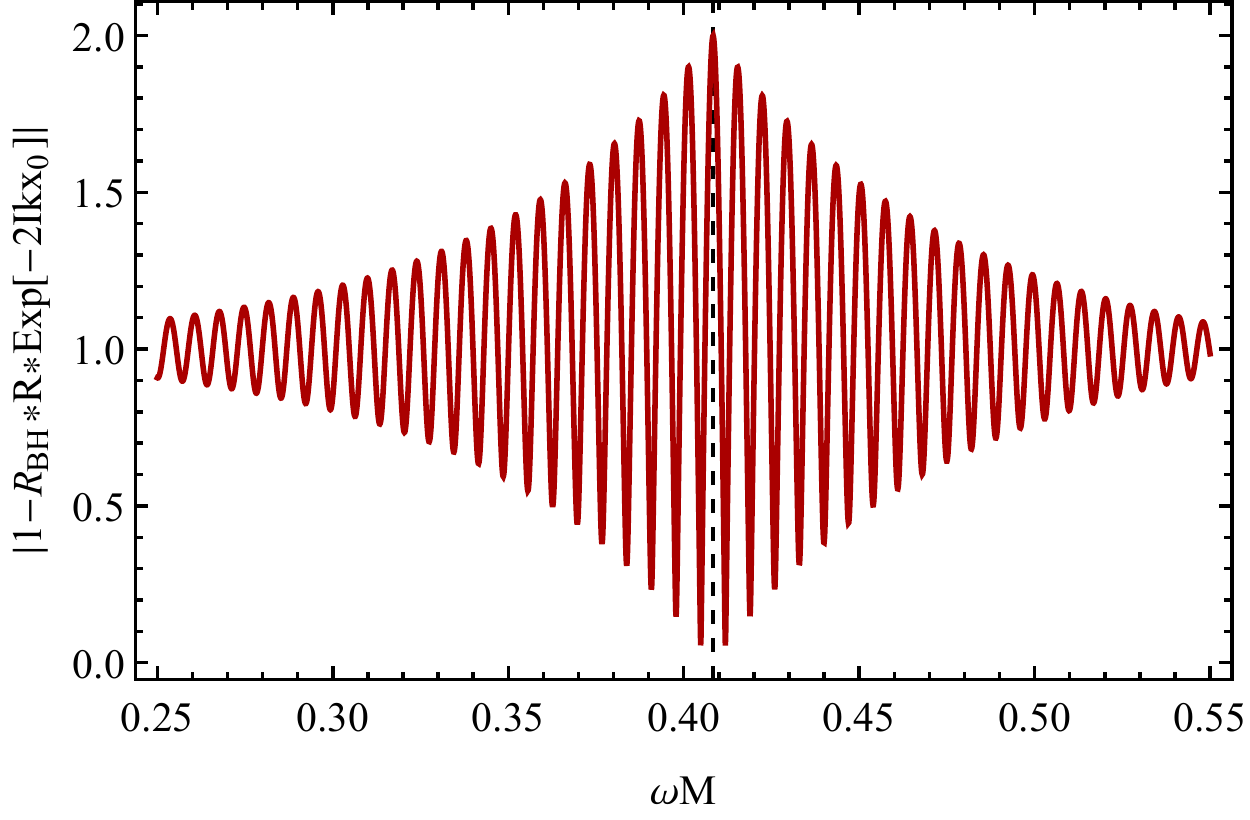}
\caption{\label{fig:AbsDenominator} The absolute value of the denominator of the transfer function $|1 - R_{\textrm{BH}}(\omega)R(\omega)e^{-2ikx_0}|$ for Boltzmann echoes. Along the real axis, the denominator never goes to zero, leading to larger imaginary parts for poles further from the superradiance frequency. Peak locations $l_r$ on the real axis are related to minima of this function.}
\end{figure}

The absolute value of the denominator of the spectrum $d(\omega) = |1 - R_{\textrm{BH}}(\omega)R(\omega)e^{-2ikx_0}|$, coming from the transfer function $K$, shown in Fig.\ref{fig:AbsDenominator}, generates the resonance structure for echoes, and through its minima provides most of the information that determines the values of $l_r$ along the real axis.  
However, modulation from the numerator, which includes the source function, still needs to be accounted for to obtain necessary accuracy (see Subsection \ref{sec:Sourcing} for discussion on source functions). 

\begin{figure}
\centering
\includegraphics[width=0.47\textwidth]{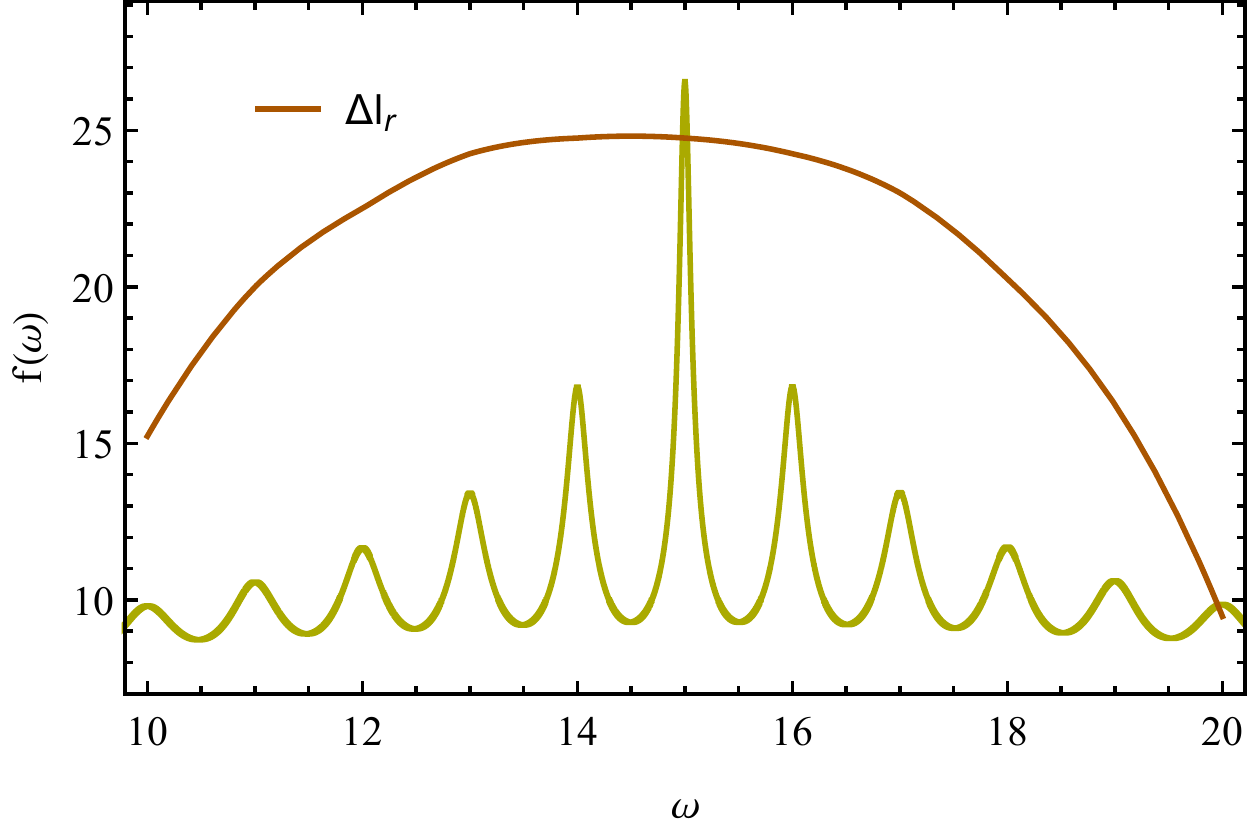}
\caption{\label{fig:OverlapShiftTest} The imaginary part in the denominator is the primary cause of the nonconstancy of the separation function, and causes peak locations of to shift away from regions of higher amplitude. Plotted here is a spectrum of simple poles $f(\omega) = \Sigma_{n=1}^{30} |\omega - n + 0.05 i (1 + 0.95 |n - 15|)|^{-1}$ representing a toy Boltzmann model, plotted with the corresponding separation function measuring the differences between peak locations. The separation function is offset above its actual value for visualization.}
\end{figure}

\begin{figure}
\centering
\includegraphics[width=0.47\textwidth]{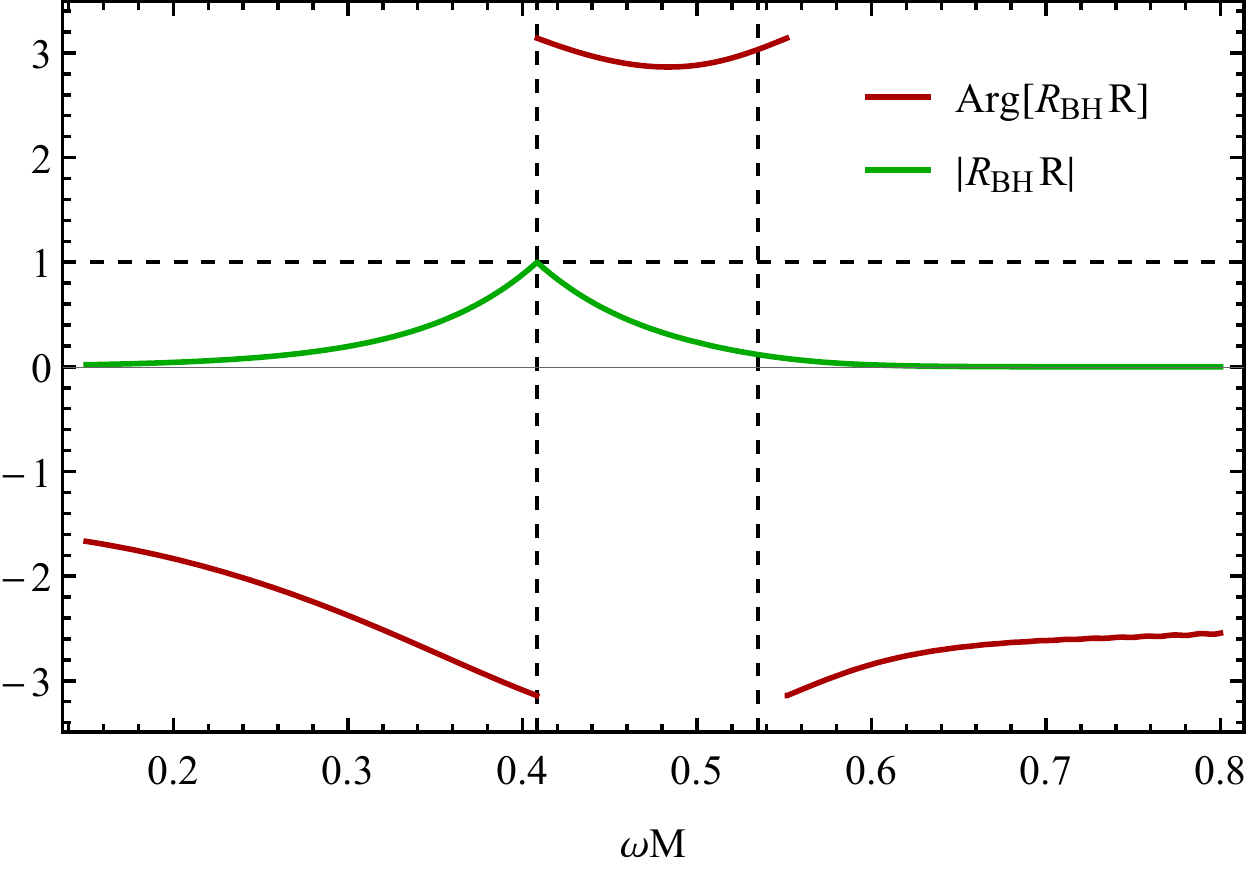}
\caption{\label{fig:RBHR} Red: The phase of $R_{\textrm{BH}}R$ for $a = 0.7$ for Boltzmann reflectivity. Green: The magnitude of the same. The horizontal dashed line is at unity, the peak of $R_{\textrm{BH}}R$. The separation function deviates from integer multiples of the inverse time delay $1/t_d \approx 1/|2x_0|$, and this deviation is dependent on the difference between the magnitude of $R_{\textrm{BH}}R$ and unity and the dynamics of the phase. Towards negative frequencies, the phase increases, causing the phase for the minimum to need to be relatively negative, implying lower frequencies and thus a stretching away from the superradiance frequency. The reverse is true towards positive frequencies.}
\end{figure}

Most of the cause of the nonconstancy of the separation function is revealed in Fig.\ref{fig:OverlapShiftTest}, where we plot the separation function for a simple sum of poles with Boltzmann-like imaginary part, providing a toy model representation of the echo spectrum.
In this plot, the true locations of the poles are exactly known, but the phase of the Lorentzians coming from the denominator shifts the $l_r$ towards the region where the imaginary part in the denominator is minimized. This information is accounted for in the model by calculating the separation of resonances from the full numerical solution. This can be explained through Fig.\ref{fig:RBHR}, where it is shown that the phase and magnitude of $R_{\textrm{BH}}R$ is not slowly varying near the superradiance frequency, and this distorts the apparent periodicity of the exponential phase factor in $P$.
The result is that $e^{-2ikx_0}$ does not minimize the denominator at constant integer multiples of the resonance frequency, but now this factor must account for phase and magnitude changes in the reflection functions. 

Measuring the peaks $l_r$ directly from the full spectrum, we numerically obtained the separation functions for each of nine sets of background parameters from the set of combinations $(a, x_0)$ where $x_0 \in \{-150, -450, -900\}$ and $a \in \{0.6, 0.7, 0.8\}$. We found two emergent patterns from which we construct a numerically efficient surrogate model. In the first, the separation functions remain approximately centered about the horizon frame frequency $k$ and are somewhat symmetric. This suggested a Taylor expansion representation in $k$, with a quartic polynomial structure yielding the approximate flattened parabolic form. At first a quadratic parabolic form was tested, but this was insufficient to model the relative flatness and asymmetry of these curves compared to parabolas. In the second, by adjusting the cavity size $x_0$, it became apparent that not only did the separation function scale linearly with $x_0$, as expected, but the curvature also decreased according to some combination of powers. We find that the Taylor expansion variable may account for this by including inverse time delay factors scaling with the cavity size. Accounting for these two patterns, we are able to write a separation function valid for all relevant background parameters as
\begin{equation}
\Delta l_r = \frac{1}{t_d}\left(c_0 + c_1 \frac{k}{t_d} + c_2 \frac{k^2}{t_d^2} + c_3 \frac{k^3}{t_d^3} + c_4 \frac{k^4}{t_d^4}\right) .
\end{equation}
Particular fits to these coefficients are given in Appendix \ref{sec:Appendix}. In Fig.\ref{fig:ModelFlow}, we show how this surrogate function flows from a state constructed at the minimal values of $a = 0.6$ and $x_0 = -150$ to the other extremal values of $a = 0.8$ and $x_0 = -900$. Although these parameter ranges were useful during testing and model building, in reality the relevant ranges are much narrower. The accuracy of this method is discussed in Subsection \ref{sec:Accuracy}.
\begin{figure}
\centering
\includegraphics[width=0.47\textwidth]{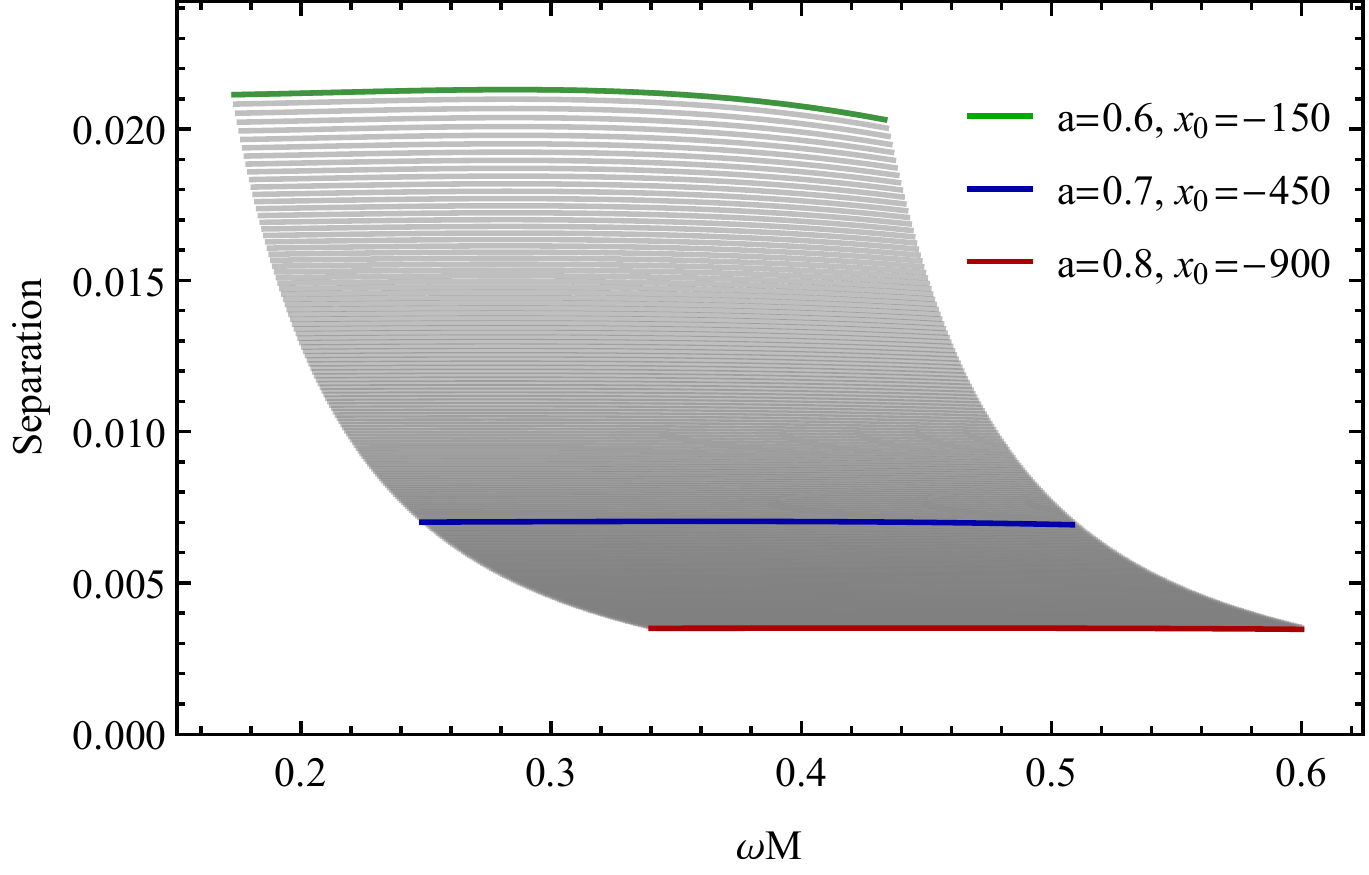}
\caption{\label{fig:ModelFlow} The modelled separation function flowing between the two extremes of tested parameter values. This function is defined by a Taylor expansion about the horizon frame frequency scaled by the inverse time delay $k / t_d$, and flattens and shifts to higher frequencies as $a$ and $x_0$ increase. The grey curves here are the model at background values between those of the green, blue, red curves which are directly calculated data sets.}
\end{figure}

\subsection{Amplitude $a_r$ and Width $w_r$}

Echo resonances are Lorentzians modulated by a frequency-dependent amplitude, and their heights along the real axis are equal to the amplitude divided by the imaginary part of the denominator. The widths of the Lorentzians are determined by this same imaginary part. The amplitudes and widths $a_r$ and $w_r$, as defined in the model above, may be measured directly with this information, for example by measuring the full heights of the resonances as well as their half-widths at the half-max points, but two issues will arise: First, the measured value of the height of a resonance is a mixture of the amplitude and the width, thus any error in the width estimation carries through into the amplitude measurement. Second, echo resonances, including those for Boltzmann echoes, are generally not sharp enough to ignore the overlap between them which forms a continuum. 

To avoid the problems caused by overlap, the envelope of overlap, defined by the portion of the curve below the line connecting minima (a.k.a. the continuum), can be subtracted to isolate resonances. This improves the height measurement by removing most of the superposition, though it also removes some of the actual resonance amplitude, but this is at the expense of accuracy in estimating the widths as this method effectively deletes the bases of the Lorentzians. Since this method targets the upper portions of each resonance, the subtracted spectrum is a sum of the caps of Lorentzians and the width is underestimated leading to relatively narrow resonances. Despite these caveats, this method  produces results consistent with the more accurate technique that will be presented shortly. However, while numerically useful, this method adds unnecessary complexity to the model by generating the need for additional parameters describing the envelope, which is not itself the physical quantity of interest. 

To calculate the amplitudes, a superior method is to consider the pole structure analytically, which is possible under the previously described geometric optics approximation. Recalling that the transfer function may be written as 
\begin{equation}
K \approx T_{\textrm{BH}}(1 + P + P^2 + ...) ,
\end{equation}
valid when $-1 < P < 1$, we demonstrated that the echo spectrum with the GR component subtracted is equal to
\begin{equation}
\psi_{\omega, echo} e^{i\omega x}\frac{T_{\textrm{BH}}P}{1 - P}S .
\end{equation}
Using this, we may simply calculate the $a_r$ as the discrete sampling of $|T_{\textrm{BH}}P S|/t_d$ at the resonance frequencies, where the $t_d$ factor comes from defining the full heights of the resonances as $a_r/w_r$, and expanding around each pole along the real axis \cite{S_Conklin_73}.

\begin{figure}
\centering
\includegraphics[width=0.47\textwidth]{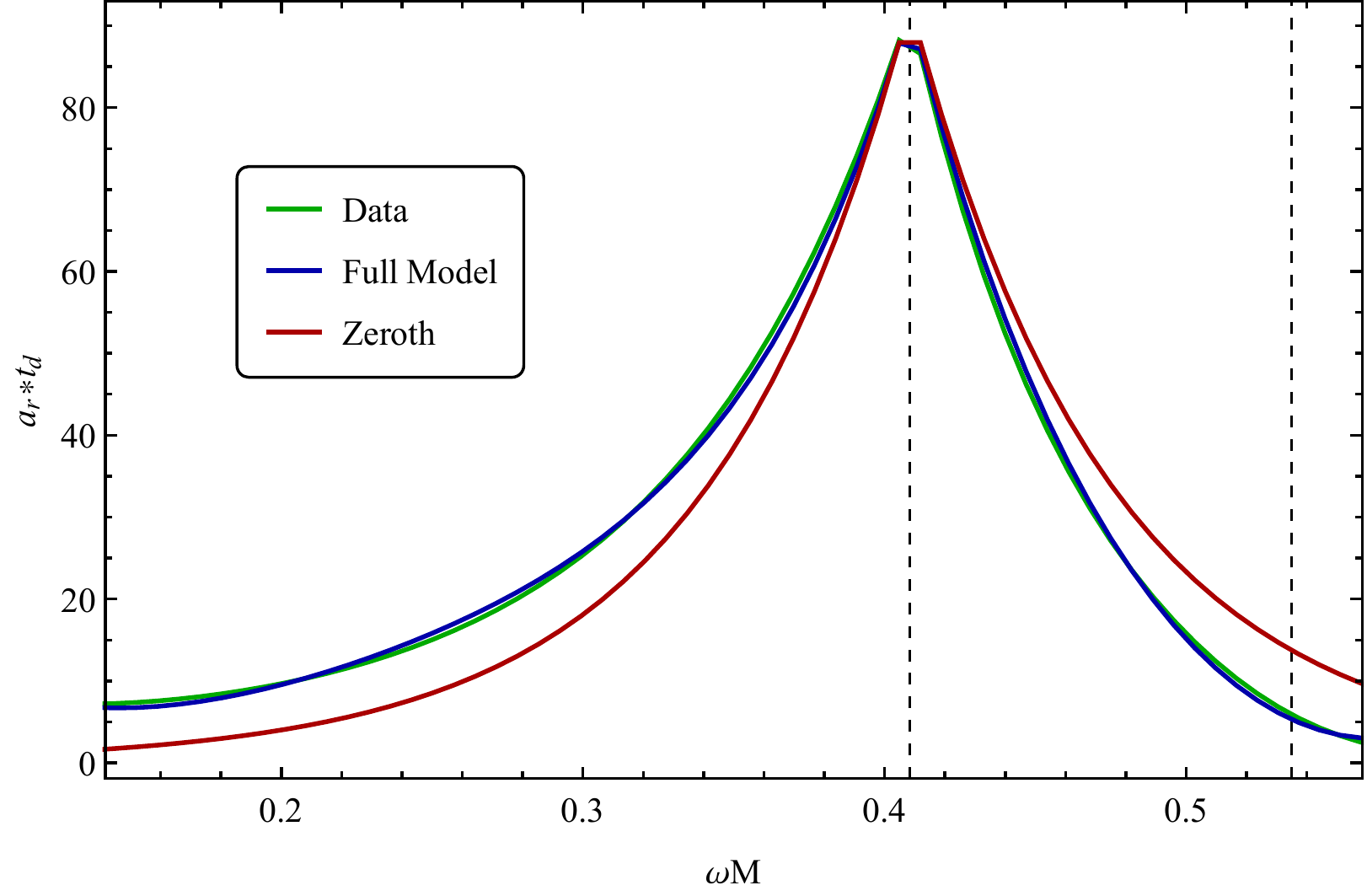}
\caption{\label{fig:ampFit} Green: The data for the amplitude multiplied by the time delay, for $a = 0.7$ and $x_0 = -450$. This data is generated from the full numerical solution to the wave equation using a GR surrogate model source. Blue: The linear model fit using the full model of $R$ plus a quartic polynomial in $k$. Red: The zeroth order fit using just $R$.}
\end{figure}

Computing the functions $T_{\textrm{BH}}$ and $P$ is expensive, despite their generally smooth form, and repeating this computation for each run of a data search would be inefficient. Rather, we create a surrogate model that closely approximates the $a_r$ and is applicable across a range of cavity sizes and background spins. This is defined by a linear fit to $R$ and a quartic polynomial in $k$ for minor corrections, and gives excellent results because of the domination of the Boltzmann factor. The best fit values for the coefficients are given in the appendix. Using the same notation as for the separation function but with different values for the coefficients:
\begin{equation}
a_r = \frac{c_{x_0}}{t_d}\left(c_R R(|k|) + c_0 + c_1 k + c_2 k^2 + c_3 k^3 + c_4 k^4\right) .
\end{equation}
Fig.\ref{fig:ampFit} gives a comparison of the data, the fit at zeroth order where just $R$ is used, and the full fit. ``Data'' here refers to the full numerically generated solution to the wave equation without the surrogate approximation.
As with the separation function, by tabulating data for the $a_r$ over the range of parameters $x_0 \in \{-150, ..., -900\}$ and $a \in \{0.6, ..., 0.8\}$, we generate a surrogate form valid for all physically relevant parameters.

The calculation of widths is similar. The half-width at half-max $\omega_w$ of a Lorentzian along the real axis is $\omega_w = \sqrt{3}|w_r|$, where $w_r$ is the imaginary part of the QNM frequency, hence the name of ``widths'' for the $w_r$. Numerically, the $w_r$ may then be measured by determining the location of the half-max, and the height may be determined as the difference between the maximum and an adjacent minimum.

Two issues must be carefully handled with this approach. First, as with the amplitudes, the actual height of the resonances is obscured to the degree that resonances are overlapping. Second, the envelope of overlap causes an imbalance between adjacent sides of each resonance and therefore an ambiguity in the definition of the min-to-max peak height, though this can be partially mitigated by averaging the heights calculated on either side if the envelope curvature is small.

\begin{figure}
\centering
\includegraphics[width=0.47\textwidth]{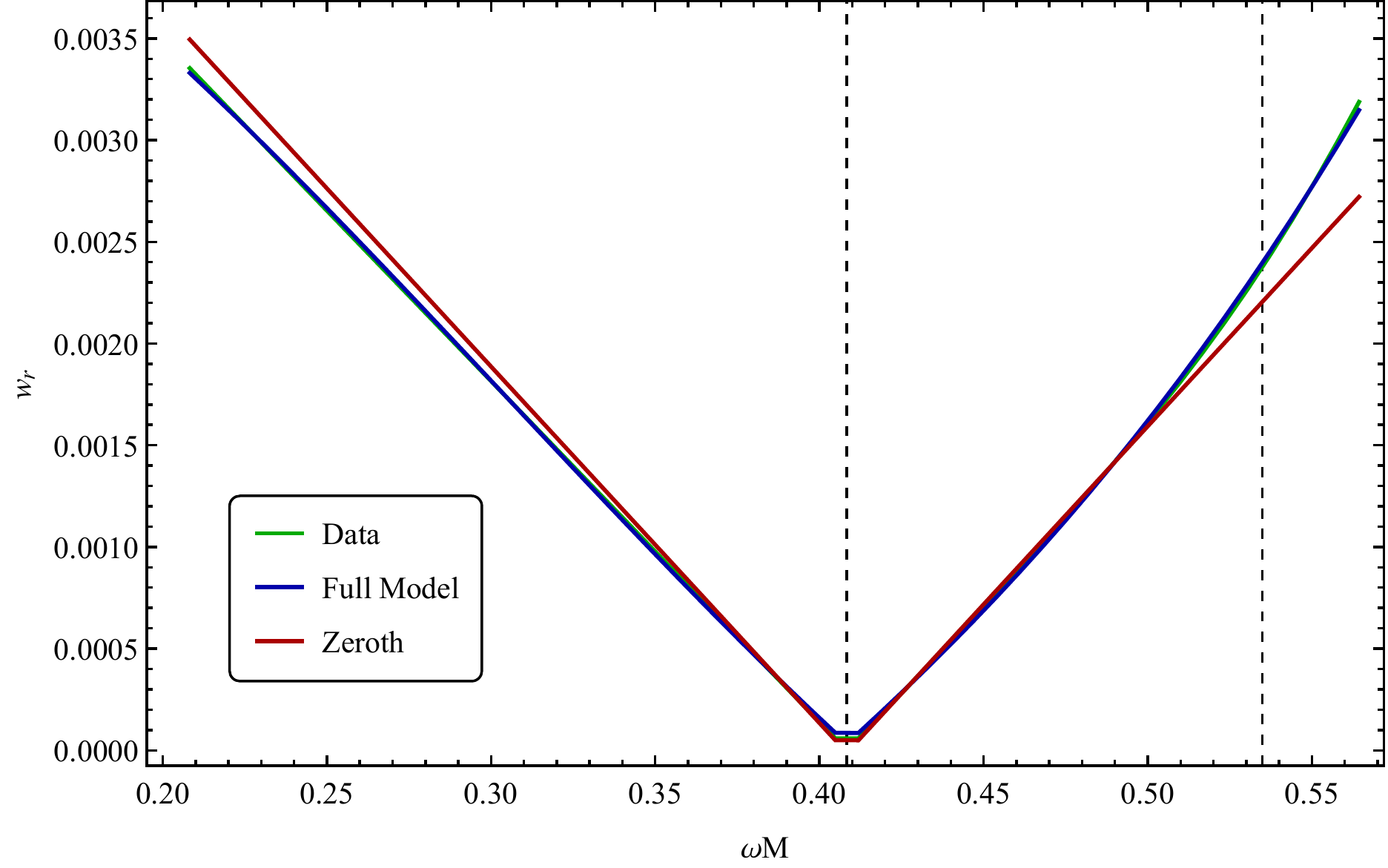}
\caption{\label{fig:widFit} Green: The data for the width parameter. Blue: The full linear model fit using $|k|$ plus a polynomial in $k$ up to third order. Red: The zeroth order approximation using just $c_0 + c_1 |k|$.}
\end{figure}

But such issues can be circumvented altogether since an approximate analytical form for the widths emerges when the boundary condition is chosen, and we find another surrogate function valid over the full range of relevant background parameters. For poles close to the real axis, and $R_{\textrm{BH}}R$ slowly varying along the imaginary axis, the imaginary part of an echo resonance is approximately equal to
\begin{equation}
\omega_r \approx \frac{ln|P|}{t_d}.
\end{equation}
The primary functional form that approximates this is $|k|$ plus an offset from the real axis. We use this and polynomial corrections in $k$ up to third order to account for nonlinearities caused by the asymmetry of $R_{\textrm{BH}}$ compared to $R$ across the axis where $k = 0$. We set
\begin{equation}
w_r = \frac{1}{t_d}(c_0 + c_1 |k| + c_2 k^2 + c_3 k^3) ,
\end{equation}
with best-fit values given in the appendix, and where these $c$'s are again distinct from those in the separation and amplitude functions. Fig.\ref{fig:widFit} compares the fits to the data for the zeroth order and full fit models.

\subsection{Phase $\phi_r$}

Here we discuss the calculation of the phase factor for each resonance, and the method used for relating this to QNM's in the time domain.

The ringdown phase of a binary inspiral occurs after the highly nonlinear merger, and is when the final object is considered to be in a perturbed final state, shedding its deformity through QNM's. These QNM's must be defined with a starting time, as they feature exponential growth in one direction (for Boltzmann echoes this is into the past) and must be truncated to prevent runaway effects. For ringdown QNM's there exists a natural starting time which is the start of the ringdown. For echoes, only the initial pulse preserves the QNM structure of the GR signal, since the QNM spectrum is different for echoes, and it may be possible that each mode of the echo QNM's is created at a different time dependent on the geometry of the cavity and the initial perturbation. 

This is plausible as the phase $\phi_r$ of each resonance causes each mode to shift in the time domain, displacing the echoes.
\begin{figure}
\centering
\includegraphics[width=0.47\textwidth]{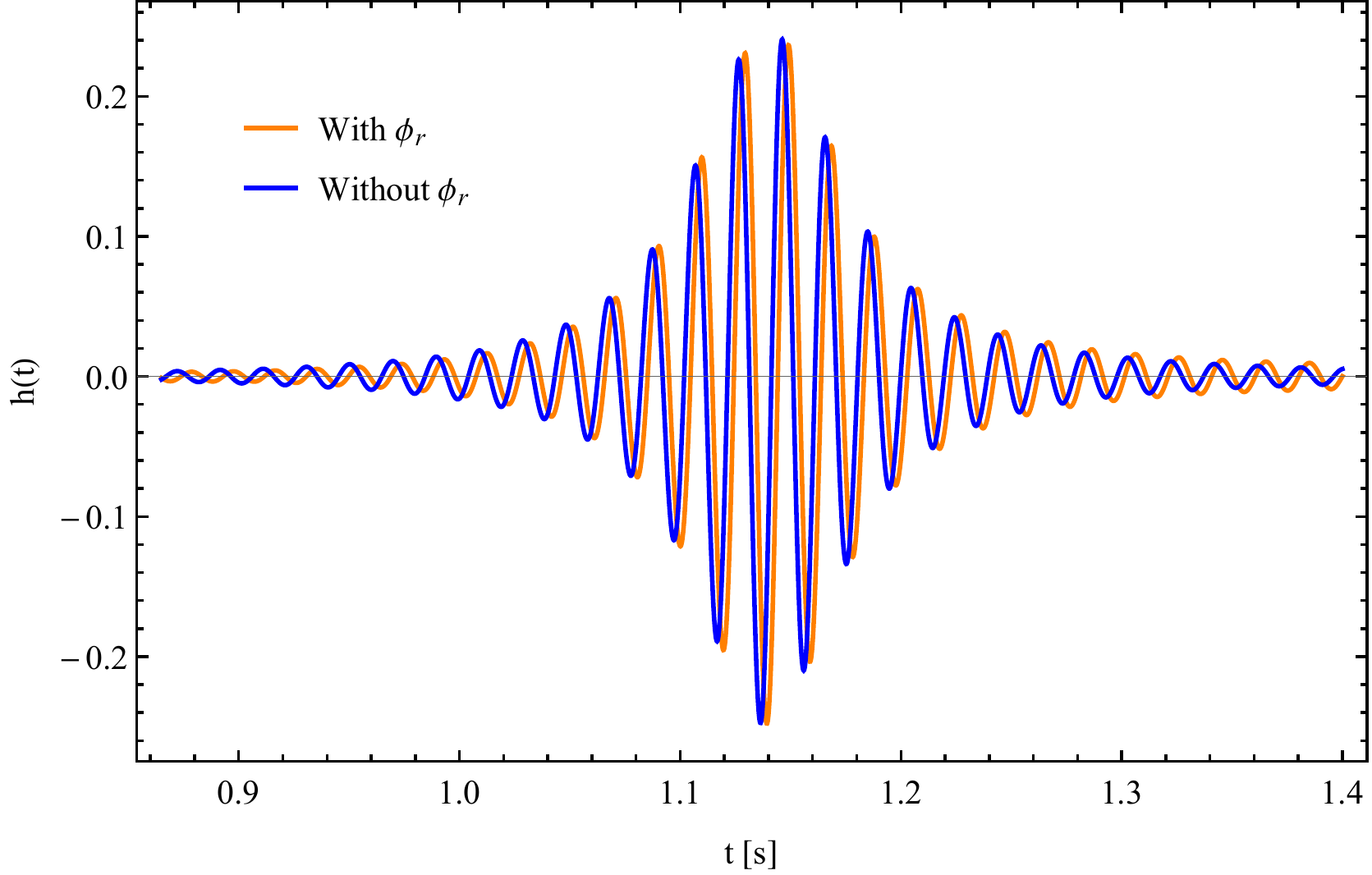}
\caption{\label{fig:EchoPhaseShift} Blue: The real part of the second echo after ringdown from the model without phase information included. Orange: The same but with phase included. The observed shift of the waveform is approximately one sixth of the wavelength for this particular echo and background with $a = 0.7$ and $x_0 = -450$. More pronounced displacements are accessible with other parameters.}
\end{figure}
Decomposing the phase for all resonances as the frequency-dependent function $\phi(\omega) \equiv \omega g(\omega)$, this time shift is equal to $-g(\omega)$. From this decomposition, and by converting to SI units for comparison with observational data, we plot a sample of the time shifting in Fig.\ref{fig:EchoPhaseShift}.

For observational considerations, it is also important to note that the measured phase is affected by the detector orientation. The full waveform $h = h_+ + i h_{\times}$ is not fully detectable by a single L-shaped detector which can only measure one orthogonal projection of this. A detector may observe instead $h_{proj} = c_+h_+ + ic_\times h_\times$ where $0 \leq |c_\pm| \leq 1$ depend on the detector orientation relative to the source.

\begin{figure}
\centering
\includegraphics[width=0.47\textwidth]{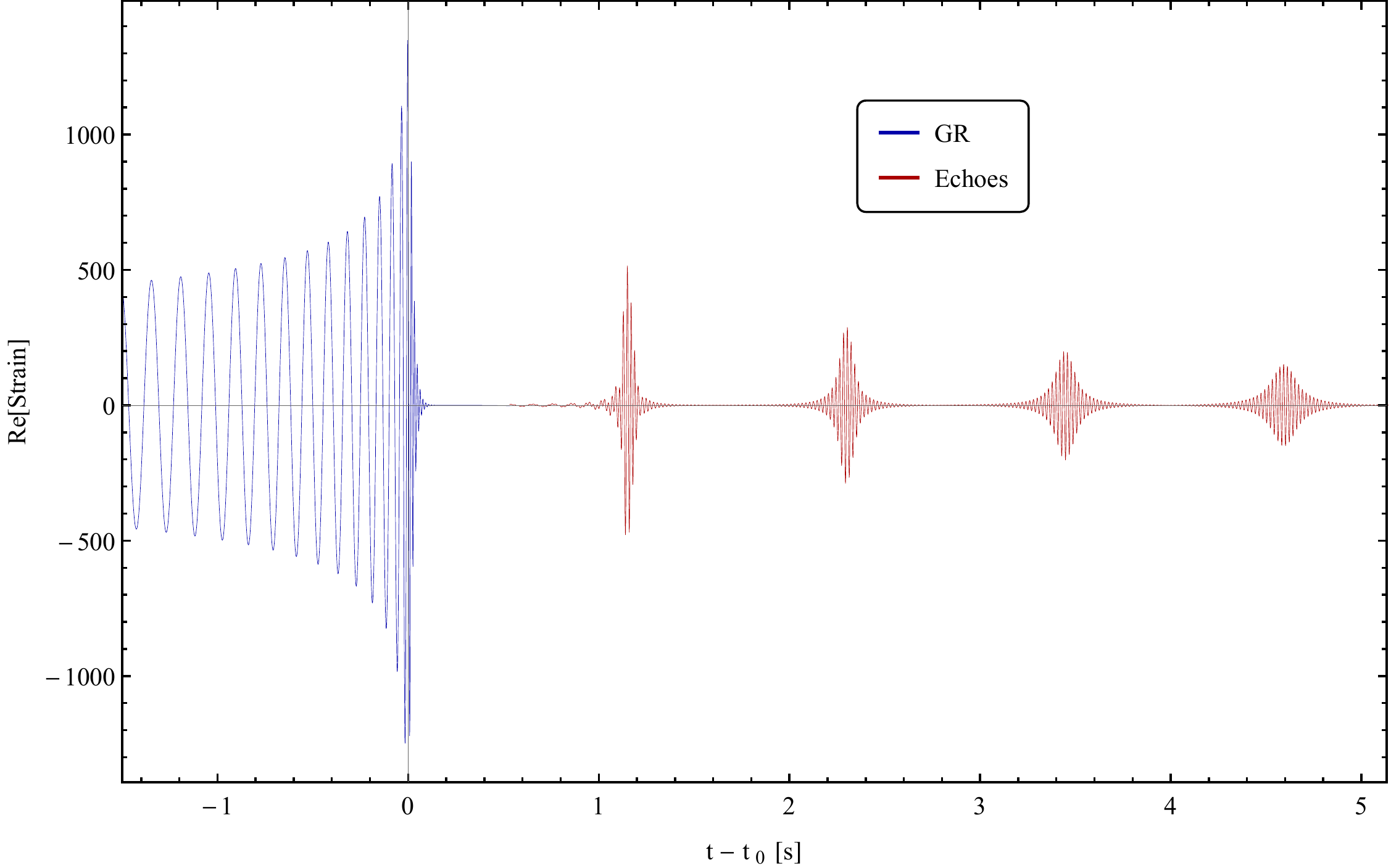}
\caption{\label{fig:psiTime} The full waveform in time reproduced by our model. Blue: The GR inspiral-merger-ringdown set to fairly closely match that observed in GW190521. Red: The echoes generated by taking this GR signal as the source. The GR component was initially generated as a numerical relativity surrogate model through the PyCBC waveform library. The parameters used to generate this are consistent with the LIGO median values and uncertainty intervals.}
\end{figure}

Accounting for all of this, we present our model in the time domain in Fig.\ref{fig:psiTime}.
Here both the GR and echo components are reproduced by our full model, with the GR waveform set to fairly closely match that observed in GW190521, which is a sample event of special interest for which we perform a data analysis in this paper.

\subsection{Accuracy}\label{sec:Accuracy}

We have now developed numerically efficient surrogate functions for each of the Boltzmann parameters. These methods are applicable for other models as well with different values for the $c$ coefficients. To test the accuracy of our surrogate model, we measure the difference between the surrogate model and the full numerical solution without approximation in two ways. First, the location of the resonances being the most sensitive parameter, we compare the modelled separation function to the data over a wide range of background parameters. Second, we test the overall accuracy of the model including all parameters by comparing the final modelled spectrum to the data for the expected physical background values.

\begin{figure}
\centering
\includegraphics[width=0.47\textwidth]{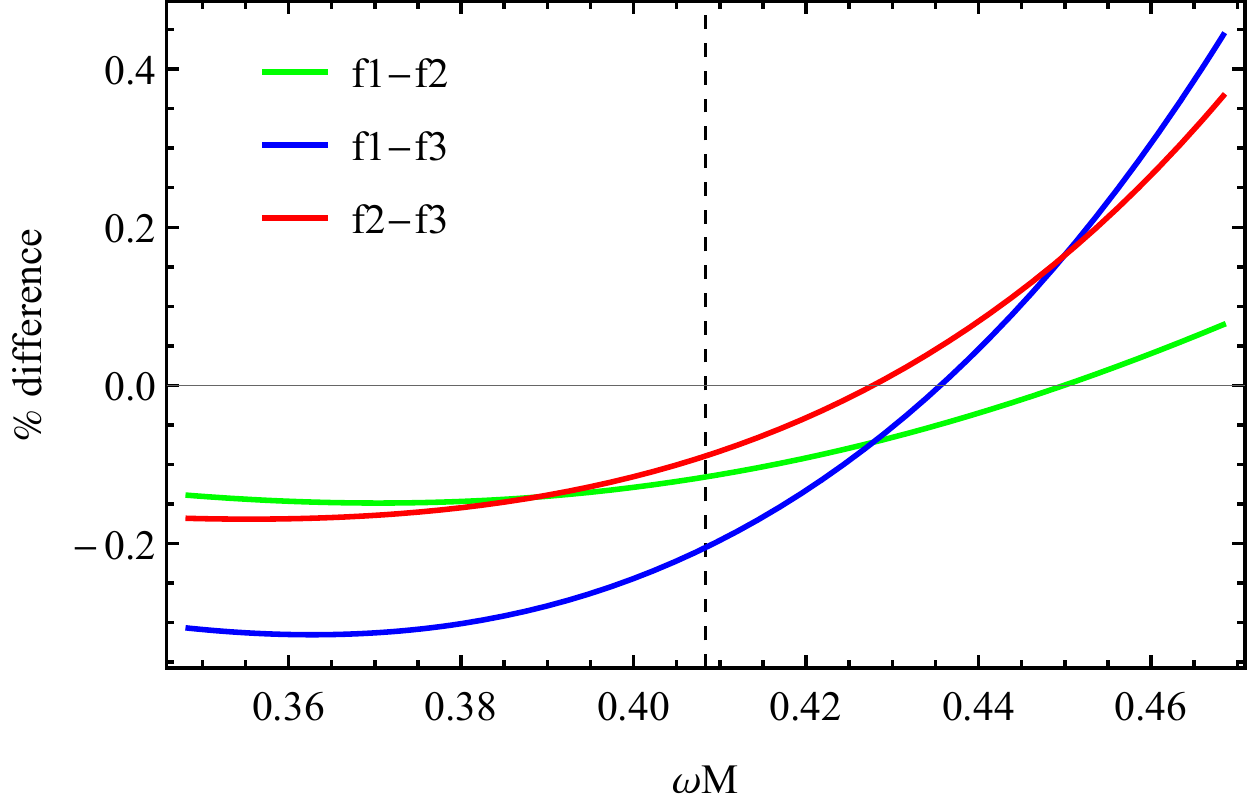}
\caption{\label{fig:PercentageDifference} Evaluating the effectiveness of the surrogate model over changes of spin. Here the surrogate model is generated directly from the data at $a = 0.6$ and $x_0 = -450$ to give $f1$, then for $a = 0.7$ and $a = 0.8$ to give $f2$ and $f3$, respectively. The colours here represent the percentage difference between the surrogate models from the directly generated fits for each spin value and the rescaled evaluations at the respective values. The parameter and frequency ranges represented here are conservative, extending through the LIGO 90\% credibility ranges.}
\end{figure}

\begin{figure}
\centering
\includegraphics[width=0.47\textwidth]{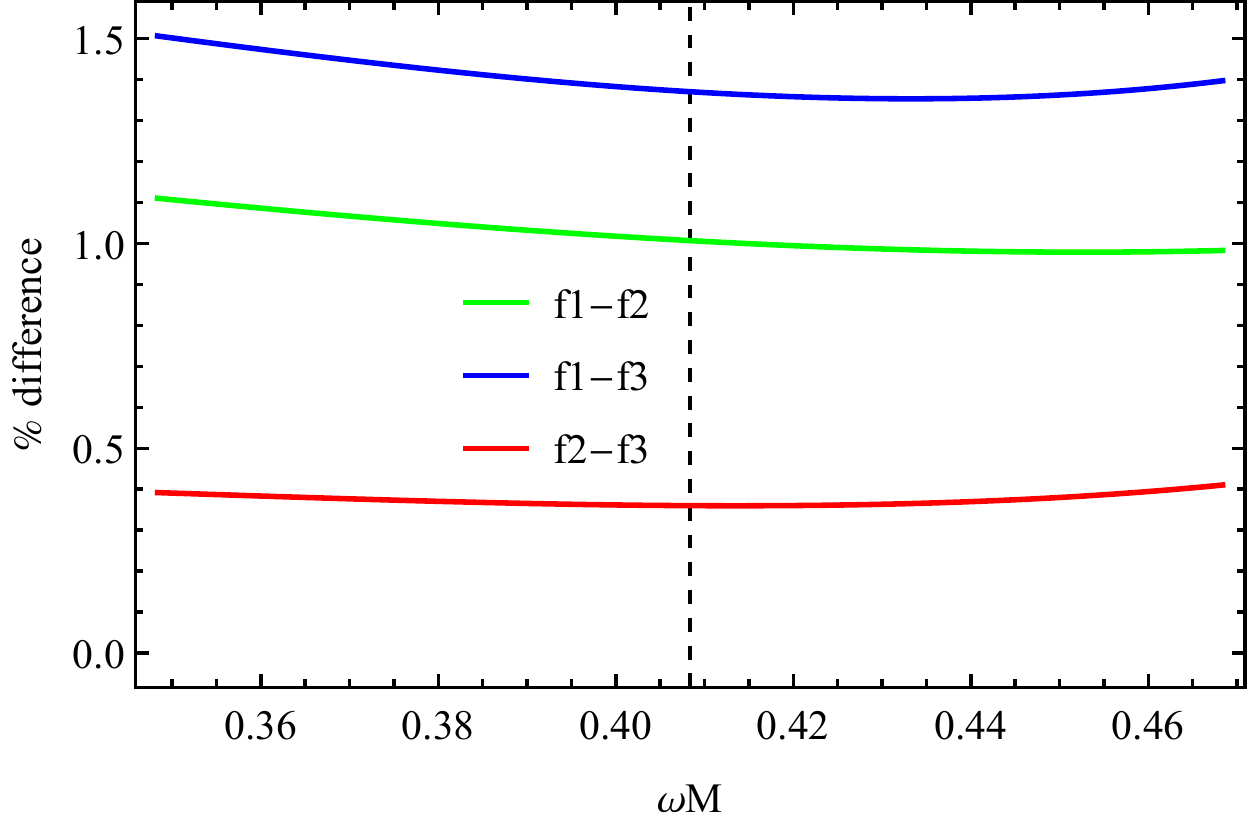}
\caption{\label{fig:PercentageDifferenceX0} Evaluating the model over changes in cavity size $x_0$. All data points here have $a = 0.7$, and $x_0$ goes from -150 to -450 to -900 for $f1$, $f2$, and $f3$, respectively. The range here is very conservative to test against edge cases for models that severely depart from Planckian deviation. Our preference for the data search is to focus on physically motivated Planckian deviations, where the accuracy of the surrogate model is good to within about $0.2\%$ or better.}
\end{figure}

To quantify the accuracy of the separation function for different spins and cavity sizes, we calculate the full spectra over extreme ranges of $a$ and $x_0$. For $a$ the range includes the values observed by LIGO, and for $x_0$ we use a very extreme range that includes Planckian deviations. For $x_0$ in particular, the actual priors are much narrower than what we present here, a choice we made to be as general as possible for different models that depart from Planckian deviation. We then use the ranges of these parameters to generate benchmark functions $f1$, $f2$, and $f3$ and compare these to the surrogate model evaluated at the same values, where the initial surrogate coefficients for the spin comparison were derived using the extremal spin value of $a = 0.6$ and for the cavity size comparison using the extremal value of $x_0 = -150$. The actual surrogate model used in the data search is developed from the more physical $a = 0.7$ and $x_0 = -450$, but we test the extremal values here to be conservative and consider edge cases. Fig.\ref{fig:PercentageDifference} plots the percentage difference between each evaluation of the surrogate function and the ``data'' for the spin comparison. For example, to obtain the blue curve $f1 - f3$, the surrogate model is generated from the data at $a = 0.6$ then rescaled to $a = 0.8$ and compared to the data at that point.
Performing the same actions across cavity sizes yields Fig.\ref{fig:PercentageDifferenceX0}. 
From these evaluations, the separation function proves accurate up to differences of between approximately 0.2-1.5\%. Practically, for a data search this error margin is much smaller since the expectation for the cavity size is set for each value of $M$ and $a$ to the approximately Planckian value and the spin is most likely closer to the best fit value rather than the extremes that we test.

\begin{figure}
\centering
\includegraphics[width=0.47\textwidth]{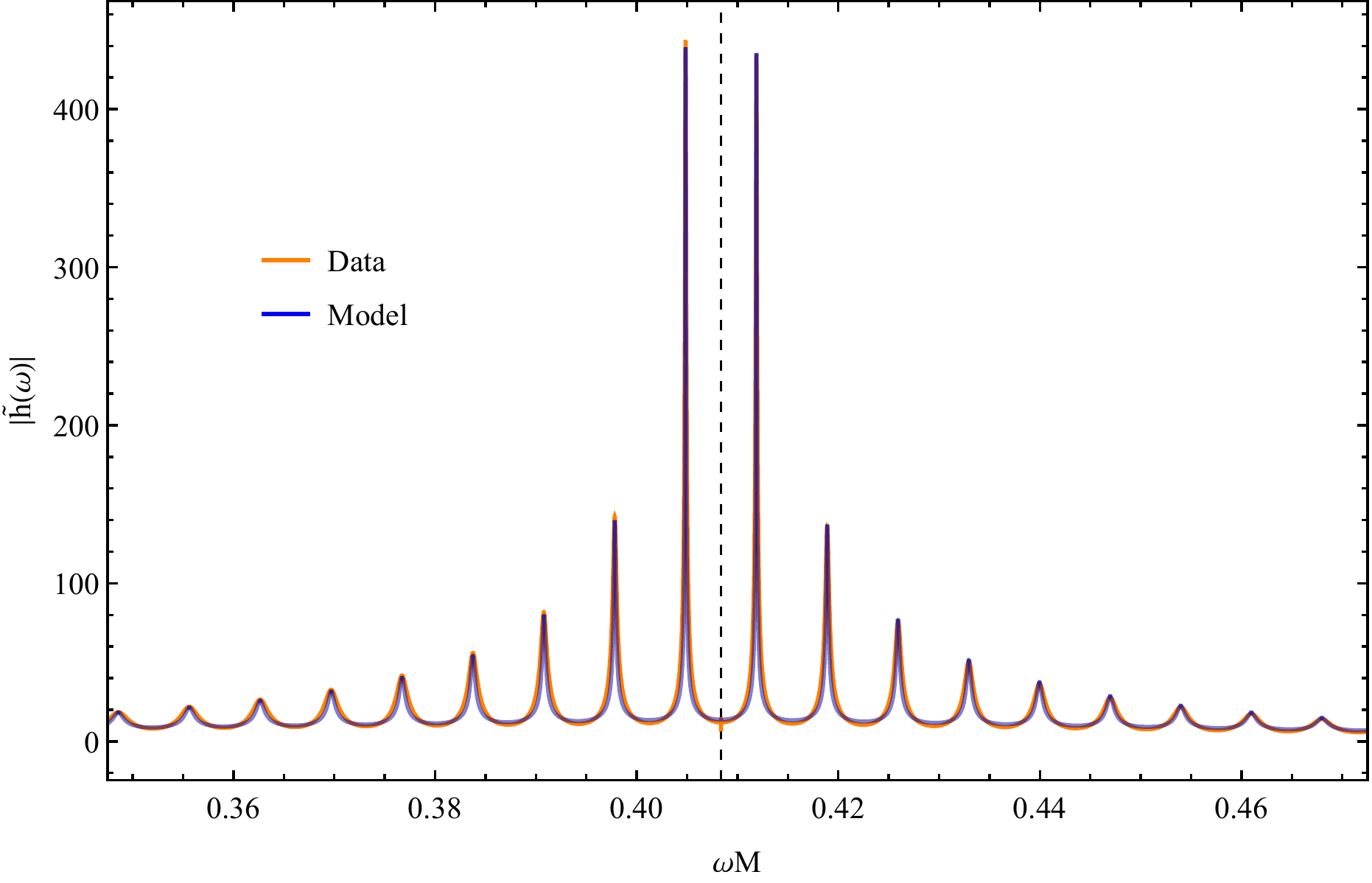}
\caption{\label{fig:ModelCompMod} Blue: The absolute value of the full surrogate model spectrum for Boltzmann echoes with $a = 0.7$ and $x_0 = -450$. Orange: The ``data'' for the same background parameters, corresponding to the full numerical solution without the surrogate approximations.}
\end{figure}

Combining all parameters, along with a final division of the width (and corresponding amplitude) by two for better agreement with the data, we plot the final form for our $\alpha = 1$ Boltzmann surrogate model using a numerical relativity motivated source in Fig.\ref{fig:ModelCompMod}, along with the data, at the LIGO-measured median background values. 

\subsection{Sourcing}\label{sec:Sourcing}

In the absence of fully nonlinear numerical GR simulations for the binary merger of ECO's, much study has been devoted to echoes sourced by simple analytical forms, such as Gaussians and delta functions. Significant progress was made in \cite{S_Conklin_24} where the connection between physical initial conditions in the Teukolsky framework and the source integral in the Sasaki-Nakamura formalism was established with applications for echoes, along with the connection to angular modes via the spin-weighted spheroidal harmonic distribution. In \cite{S_Conklin_38}, further progress was made in demonstrating that a random distribution of Gaussians superimposed to create a somewhat amorphous initial perturbation could still lead to distinct resonances, albeit with expected distortion. However, these works did not yet clarify what the physical source distribution \textit{should} be, but rather what it \textit{would} be under certain reasonable conditions. This task of physically motivating a source was partially accomplished for a static background in \cite{S_Mark_79} which used a toy model simulation of a point particle falling into a static background along an ``ISCO'' orbit, and collected the source information numerically. This provides a helpful comparison to other work as this more physical source was largely fit by a Gaussian, but with higher order structure and width and location defined by the toy model inspiral. Developing this to a more relevant scenario, \cite{S_Micchi_91} upgraded the toy model to include spin. The results were qualitatively similar, and affirmed progress made in earlier work. 

\begin{figure}
\centering
\includegraphics[width=0.47\textwidth]{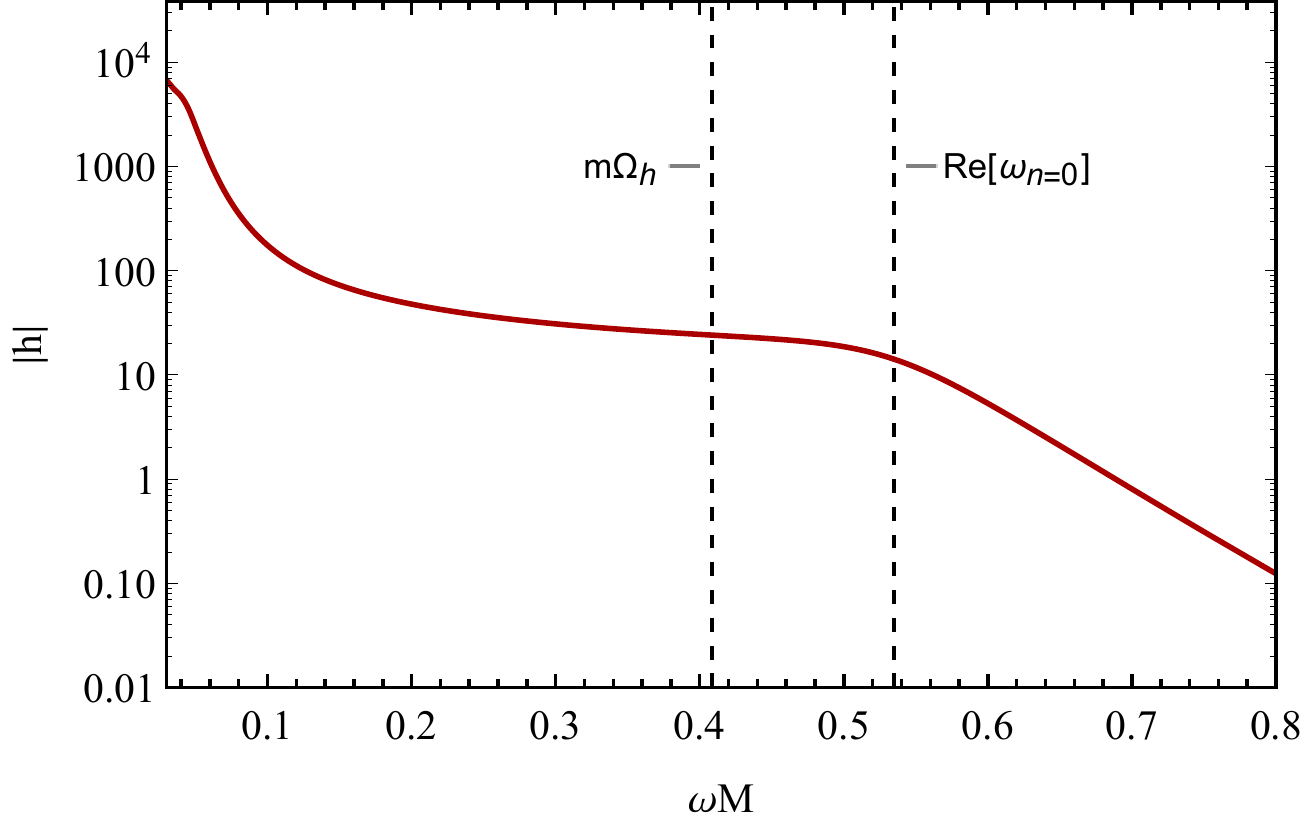}
\caption{\label{fig:Surrogate} The PyCBC surrogate model from the SEOBNRv4\_opt approximant with $mass1 = 40.0$, $mass2 = 40.0$, $spin1z = 0.09$, and $spin2 = 0$, sampled at 4096Hz with $distance = 1280Mpc$. With these parameters, and by the SEOBNRv4 approximant, the final remnant is calculated to have $a = 0.7$ and $M = 76$. The mass and distance scales of the surrogate models are irrelevant for model building and data analysis since we construct the model with the dimensionless quantity $\omega M$ and the normalization of the model cancels out in the data search.}
\end{figure}

In this paper, we gratefully make use of the surrogate model GR waveforms freely available through PyCBC \cite{S_PyCBC_d}. Since our surrogate model is developed through interpolation between nine sets of background parameters, for each combination of $(a, x_0)$ values with $x_0 \in \{-150, -450, -900\}$ and $a \in \{0.6, 0.7, 0.8\}$, we required the use of one surrogate waveform for each spin (GR surrogate models are independent of $x_0$). Using the SEOBNRv4 approximant, we took binary merger parameters of $mass1 = 42.1$, $mass2 = 32.7$, and $spins = 0$ to obtain $a = 0.6$; $mass1 = 40.0$, $mass2 = 40.0$, and $spin1z = 0.09$ to obtain $a = 0.7$; and $mass1 = 40.0$, $mass2 = 40.0$, $spin1z = 0.7$, and $spin2z = 0.075$ to obtain $a = 0.8$. For the model-building and analysis, we converted to dimensionless frequencies $\omega M$ so that our results are independent of the mass scale and the total masses for each of the surrogate models is arbitrary. Because the normalization of the model cancels in the data search, the distance, which we chose to be 1280Mpc, is also arbitrary. The most important features of these waveforms in frequency space, see Fig.\ref{fig:Surrogate}, are the elbow near the fundamental QNM frequency at $\omega_f \approx 1.5251 - 1.1568 (1 - a)^{0.1292}$ \cite{S_Berti_j} and the adjacent plateau at lower frequencies, since the Boltzmann reflection exponentially suppresses the spectrum away from the superradiance frequency.
To get the frequency domain waveform we apply a Tukey window with $\alpha = 1/8$ (not to be comfused with the Boltzmann temperature variable of the same name) in the time domain and Fourier transform.

\begin{figure}
\centering
\includegraphics[width=0.47\textwidth]{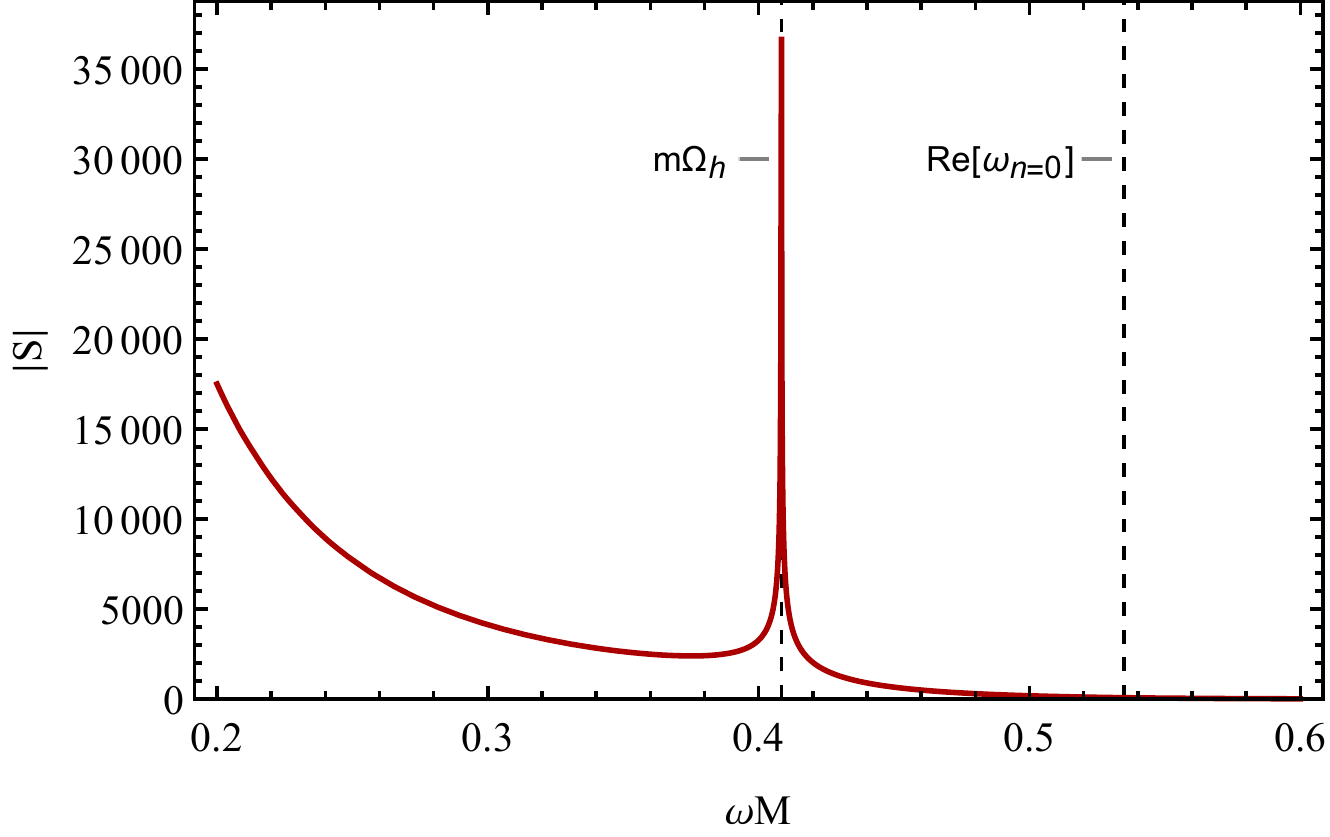}
\caption{\label{fig:Source} The absolute value of the source function $S$ which reproduces the standard GR waveform when $R = 0$, obtained through a PyCBC surrogate model. Because of the necessary division by the initial pulse factor from the transfer function $K$, $|S|$ grows very large at low frequencies, and is singular at the superradiance frequency. See Fig.\ref{fig:RBHTBH} for the  transmission coefficient from the transfer function which serves as the divisor.}
\end{figure}

The first equation of this paper references what we call the source function $S$. To ensure that our results reproduce the standard GR waveform when the background object is a standard black hole with no exotic boundary condition, $S$ must be divided by the initial pulse term from the transfer function. Specifically, we must define
\begin{equation}
S = -\frac{c_0}{4} \frac{h(\omega)}{T_\textrm{BH}(\omega)} e^{-i\omega x_0} ,
\end{equation}
where $h$ is the GR waveform. We plot this $S$ in Fig.\ref{fig:Source}.
It is important to note that for Boltzmann echoes, this form of the source function is accurate because there is exponential suppression around the horizon frequency. However, for models with boundary conditions that have less suppression at lower frequencies, an inconsistency arises. The inspiral and merger components of the GR waveform are too large to be included in the linear perturbation theory discussed in this paper, thus necessitating either a modified source function or some bandpass that removes the inspiral and merger frequencies. To use the same motivation for the source function as presented here, the echo analyst may therefore need to consider an additional truncation.

\section{GW190521}\label{sec:GW190521}

The highest network SNR gravitational wave event yet detected by LIGO and Virgo is the binary neutron star merger GW170817, with an observable signal lasting approximately 100s for an SNR of about 33.0. The next loudest are the far more massive binary black hole merger events GW150914 and GW190521\_074359 with network SNR's of about 24.4 each. While these events may be promising for the detection of echoes, as has been suggested by some searches \cite{S_Abedi_18}, we select another event with less network SNR (14.4) but significant ringdown, GW190521, not to be mistaken with the previously mentioned GW190521\_074359 occuring about five hours later \cite{S_Abbott_e}.

GW190521 is unique in several ways. It is by far the most massive event yet recorded, leading to a median measurement of the final mass of 142$M_\odot$. The observable signal is also relatively very short lived, lasting only about 0.1s and seen in the frequency range 30-80Hz with peak amplitude at 60Hz. Notably, this peak frequency is the same as the U.S. mains power signature, hence the importance of noise subtraction and independent detector correlation. The primary reason why we are interested in this event is that it may be especially significant for echoes since it generates a relatively large SNR in a short duration of time, suggesting strong ringdown, and it is the ringdown portion of the waveform that sources the echoes.

In this section we discuss our processing methods for the LIGO data, the application of our model in the search method, and our characterization of the signal that we see along with a statistical analysis of the results. In summary, we find a signal with moderate statistical significance, and we interpret this as a measurement of the energy found in the echoes by relating the observation to simulated injections on background data sets.

\subsection{Processing the Data}

LIGO data is affected by instrumental and environmental noise, and techniques have been developed for quantifying this and cleaning the data. A key quantity here is the power spectral density (PSD), defined as follows. To calculate the PSD by Welch's method, first divide a relatively large data set surrounding the region of interest into a number of segments of duration $T$, then smoothly window each segment. This windowing is done to prevent spectral artifacts that would otherwise result from the sharp truncation (square windowing) of each data segment. Then Fourier transform each segment, take the square magnitude at each frequency, and average the results to obtain the PSD. Here the important factors are the time $T$, the windowing method, and the total duration of the data from which all segments are taken. $T$ sets the frequency resolution and determines the number of segments that will be averaged, thereby setting the noise reduction at each frequency. We use Tukey windows that are tapered cosines at the edges with flat tops. These are equivalent to Hann windows when the plateau region is set to have $0$ width. We choose the plateau region to cover half of each segment, where each segment is 4s long, consistent with LIGO recommendations. We also select a $1/8$ overlap between segments. We use 1024s of data centered around the main event.

\begin{figure}
\centering
\includegraphics[width=0.47\textwidth]{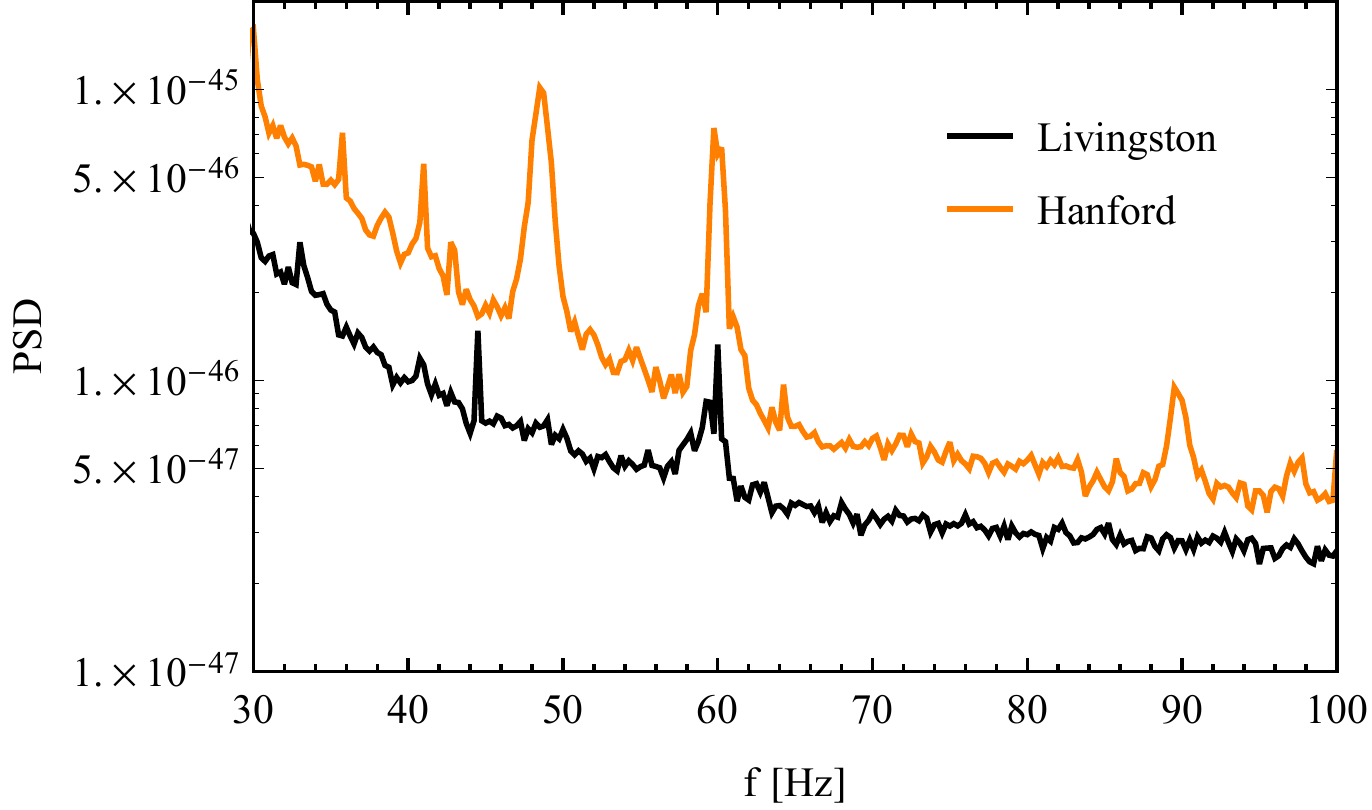}
\caption{\label{fig:PSDs} The power spectral densities (PSD's) computed by Welch's method across the frequency band relevant for echoes from GW190521. The peak at 60Hz in both detectors is at the frequency of the U.S. mains power signature. The amplitude spectral density (ASD) is the square root of the PSD.}
\end{figure}

This PSD is then used in the process of whitening the data by interpolating between the points of the PSD and dividing the spectrum of the desired data by this quantity. These interpolated PSD's for the Hanford and Livingston detectors are shown across the relevant frequency band in Fig.\ref{fig:PSDs}.
In addition to dividing out noise by the PSD, a bandpass is also imposed to cut out very low frequencies (below about $\sim$ 20Hz) that are known to be dominated by noise and approaching the uncalibrated region. Higher frequencies are also cut to avoid Nyquist sampling ambiguity and to filter out noise beyond the region of interest. We choose a somewhat tight bandpass of 30-100Hz as this satisfies noise cancellation requirements while preserving the echo signal with some buffer to account for errors in final remnant parameter estimates by LIGO, namely the mass $M = 142_{-16}^{+28}$ and spin $a = 0.72_{-0.12}^{+0.09}$, where the errors correspond to the 90\% credible intervals that include statistical errors \cite{S_Abbott_e}. To accomplish this, we use the recommended Butterworth filter which has a steeper cutoff at lower frequencies and softer taper at higher frequencies, with a flat plateau in between.

After whitening and bandpassing, the inverse Fourier transform is taken to put the data back into the time domain. At this stage a Tukey window is placed over the desired data segment and the data is ready for analysis. We use the same shape of Tukey window as previously described, but to search for echoes we choose $T = 15s$ to increase the resolution and enhance the sharp resonances. For the best fit parameters, this corresponds to about 14 echoes after redshifting the signal to the detector frame. When we need to refer to a specific value of redshift, such as when approximating the number of echoes within $T = 15s$ just now, we assume the median fit redshift of $Z = 0.82$.

\begin{figure}
\centering
\includegraphics[width=0.47\textwidth]{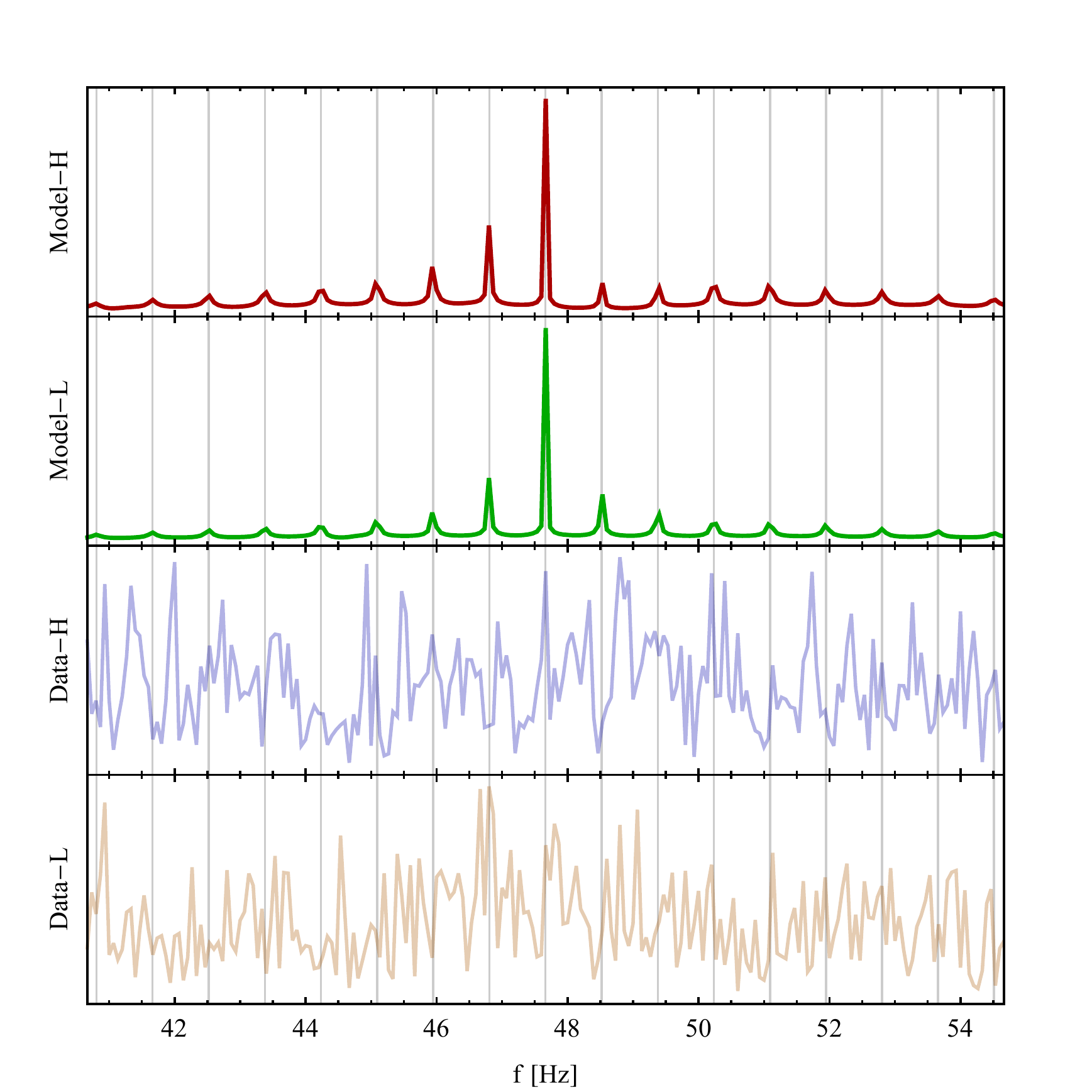}
\caption{\label{fig:BFExample2} The standard Boltzmann model with $\alpha = 1$ at the final remnant parameters $M$ and $a$ that gives the maximum SNR, plotted alongside the data, for each of the two LIGO detectors (H = Hanford and L = Livingston). Here $T = 15s$, and the frequency range is symmetric about the superradiance frequency. For visualization, the absolute value is taken for each set of data. Each curve in this figure is whitened and bandpassed consistent with LIGO recommendations.}
\end{figure}

\begin{figure}
\centering
\includegraphics[width=0.47\textwidth]{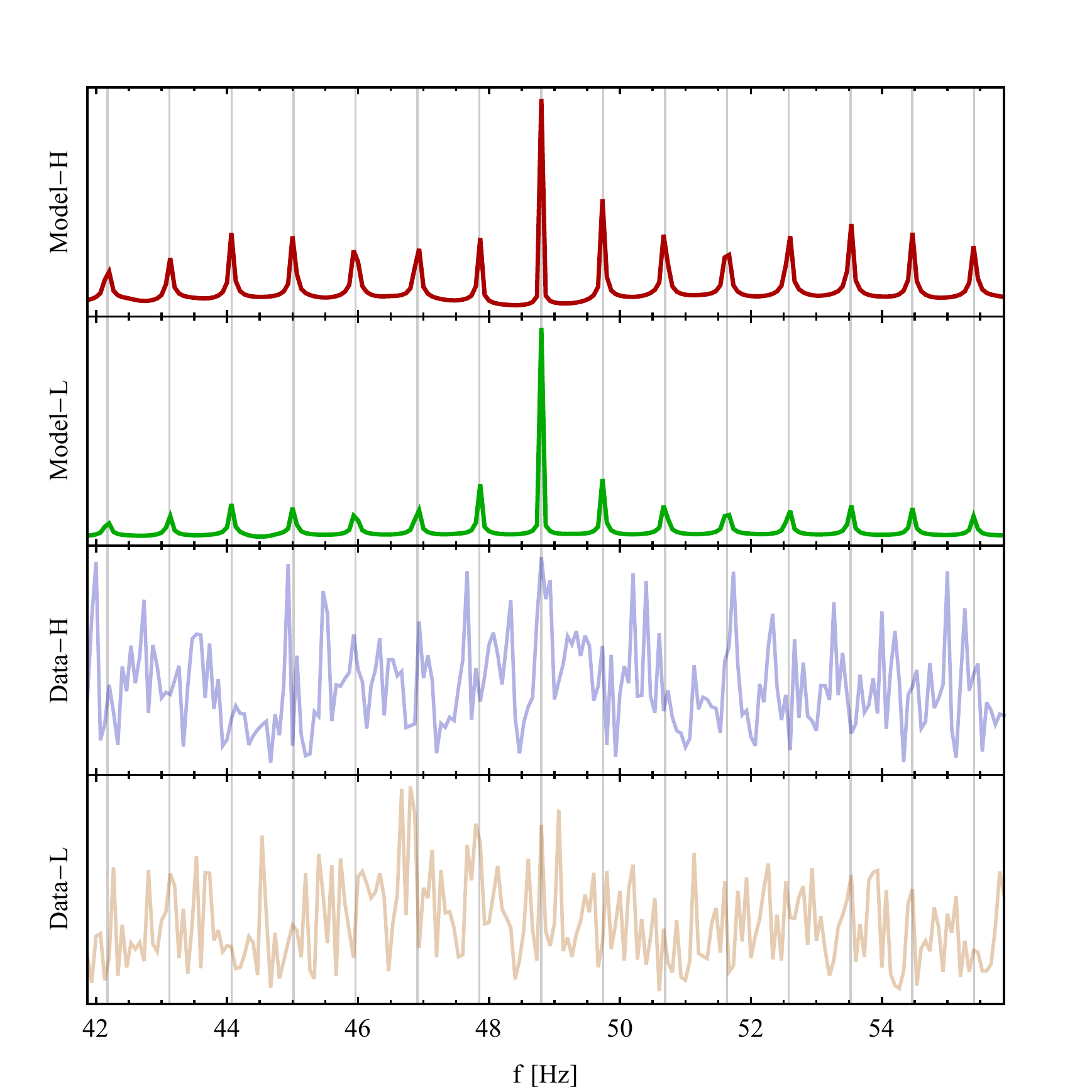}
\caption{\label{fig:BFExample2a2} The Boltzmann model with $\alpha = 2$ at the final remnant parameters $M$ and $a$ that gives the maximum SNR, plotted alongside the data, for each of the two LIGO detectors (H = Hanford and L = Livingston). Here $T = 15s$, and the frequency range is symmetric about the superradiance frequency. For visualization, the absolute value is taken for each set of data. Each curve in this figure is whitened and bandpassed consistent with LIGO recommendations.}
\end{figure}

At last, since the Boltzmann model is developed in Fourier space, we Fourier transform the data and directly compare it to the echo spectrum. Doing this consistently requires that the model itself undergoes parallel processes of whitening and bandpassing. Since our model is scalable and tested through ranges of spin from 0.6 to 0.8, and mass simply changes the frequency scale, we write our model as a function of spin and mass and calculate the overlap between the data and the model for each. To get our model as a function of just the background parameters of $a$ and $M$, we do a dimensional reduction by assuming $x_0 = -450$, corresponding to nearly Planckian deviations. For exactly Planckian deviations, $x_0$ depends on the spin and logarithmically on the mass of the background (see Equation \ref{eq:TimeDelay}), and we rely on the degeneracy that arises between the fixed cavity size model at different spin and mass values and the exactly Planckian cavity size model to accurately capture the echo waveform. This degeneracy occurs since $a$ determines the location (or shift) of the resonances and $M$ determines the scaling of frequencies. Example plots of the model overlaid with the data for each detector are shown in Fig.\ref{fig:BFExample2} and Fig.\ref{fig:BFExample2a2}.

\subsection{Analysis}

With the data cleaned and the parametrized surrogate model ready for application, we begin to look for echoes by calculating the combined signal-to-noise ratio (SNR) for the Boltzmann model in the two LIGO detectors as
\begin{equation}
SNR = \frac{1}{\sqrt{2}N_f}\frac{|\Sigma_f d_{H,f} m_{H,f}^*| + |\Sigma_f d_{L,f} m_{L,f}^*|}{\sqrt{\Sigma_f (m_{H,f}m_{H,f}^* + m_{L,f}m_{L,f}^*)}} ,
\end{equation}
where $N_f$ is the length of the data sets $d_{H,L}$ and $m_{H,L}$. Here, we have maximized SNR over an arbitrary phase difference between the Hanford and Livingston detectors. We limit the frequency range to a symmetric interval about the superradiance frequency with a half width of 7Hz, since beyond this the auxiliary resonances become less significant. The superradiance frequency is a function of spin, mass, and redshift, so the location of this interval depends on the assumed parameters for each calculation. This calculation is performed over mass and spin values derived from Monte Carlo chains that account for the probability of each set of parameters being realized in the data.

Using this SNR equation as a search tool, our method is as follows. 

\begin{figure}
\centering
\includegraphics[width=0.47\textwidth]{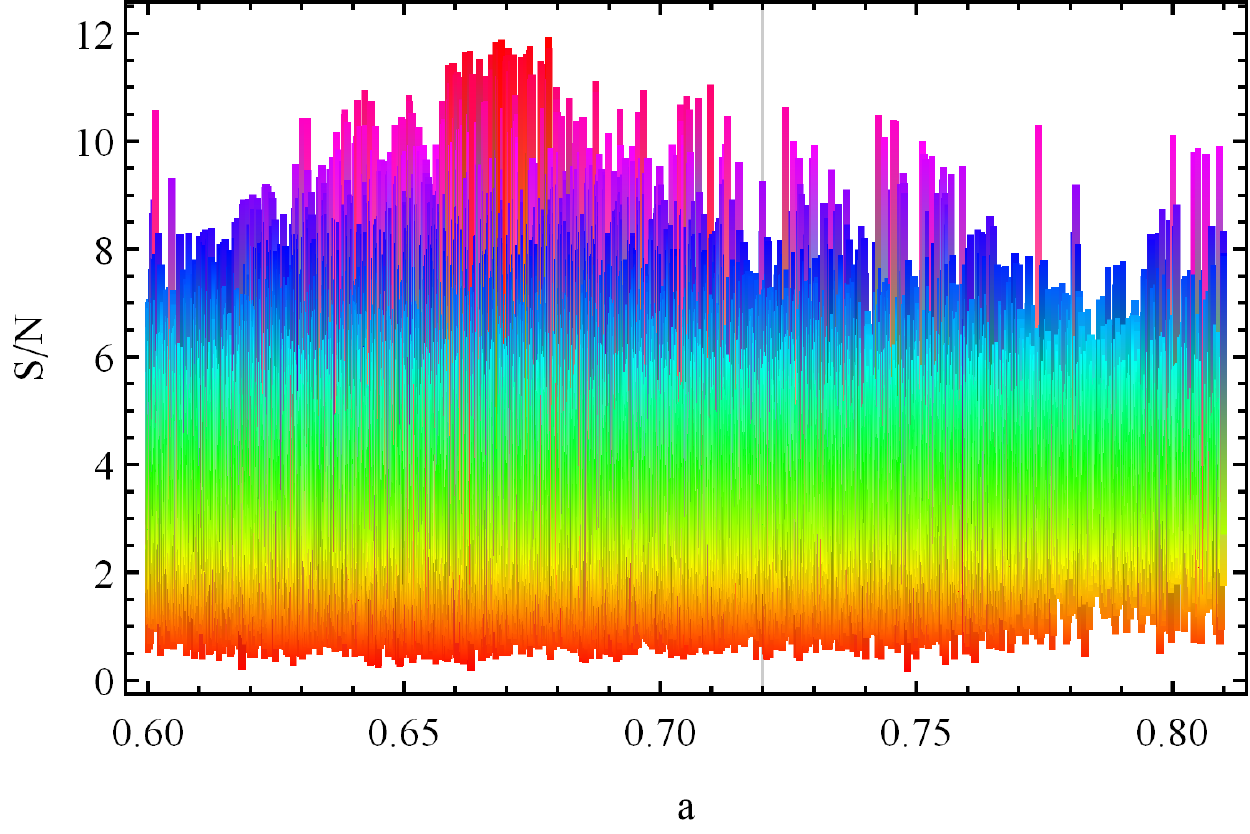}
\caption{\label{fig:EchoDatSearchBoltzA1} SNR as a function of spin for a data search with over $10^5$ samples. Each value of spin is paired with a redshifted mass value. The maximum value here is approximately 12. This search was done using the standard Boltzmann model with a 15s time segment starting at a GPS time of 0s compared to the LIGO event time. Because of the time-domain windowing, the GR event is effectively deleted from the data and does not interfere with these results. The grey line in the background marks the best fit spin.}
\end{figure}

\begin{figure}
\centering
\includegraphics[width=0.47\textwidth]{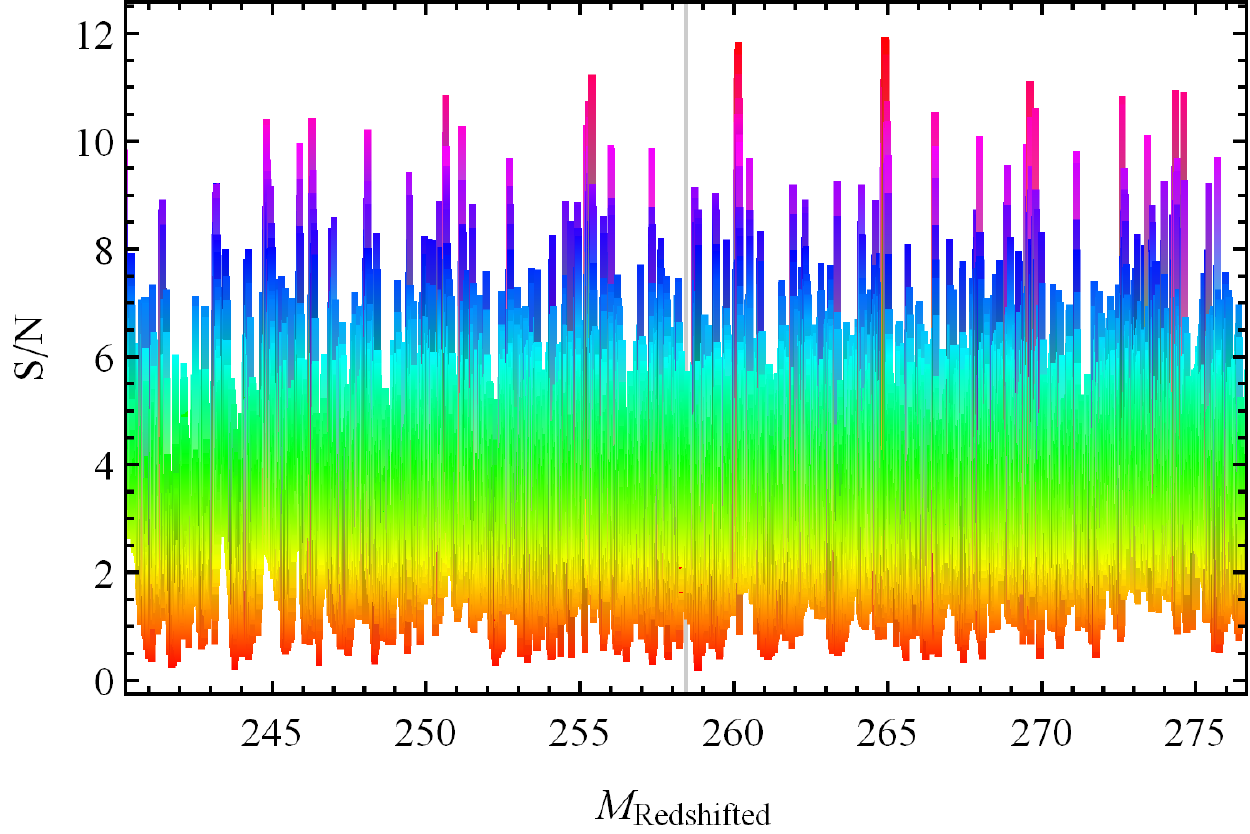}
\caption{\label{fig:EchoDatSearchBoltzA1Mass} The SNR for the same data search as in Fig.\ref{fig:EchoDatSearchBoltzA1}, but ordered as a function of the redshifted mass. Several peaks are observed at somewhat even intervals. The grey line in the background marks the best fit mass.}
\end{figure}

First, we search the data immediately after the event, where echoes would exist, and tabulate a set of SNRs for each set of $(M, a)$ (see Fig.\ref{fig:EchoDatSearchBoltzA1} and Fig.\ref{fig:EchoDatSearchBoltzA1Mass}). 
In doing so, we find a high SNR signal at a mass and spin well within the 90\% credibility ranges for the background parameters quoted by LIGO. 

\begin{figure}
\centering
\includegraphics[width=0.47\textwidth]{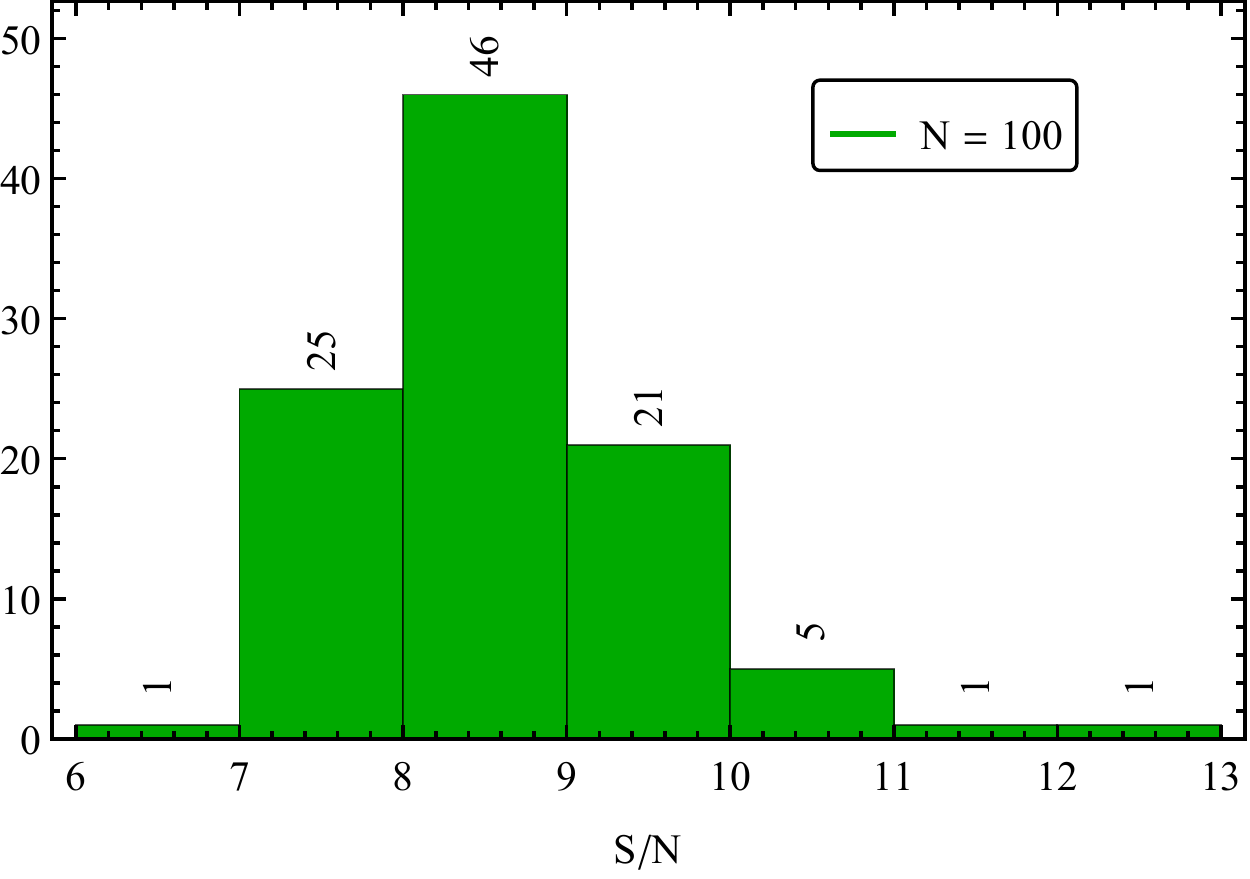}
\caption{\label{fig:EchoBackHist} A histogram of the maximum SNR found within each background data set for the standard $\alpha = 1$ Boltzmann model, showing a strong bias towards SNRs between 8 and 9, with a quick falloff towards higher values. This suggests a most probable measurement of SNR for a random background of between 8 and 9. The measured signal SNR is just under 12, so this plot suggests a p-value of about 1 in 50.}
\end{figure}

\begin{figure}
\centering
\includegraphics[width=0.47\textwidth]{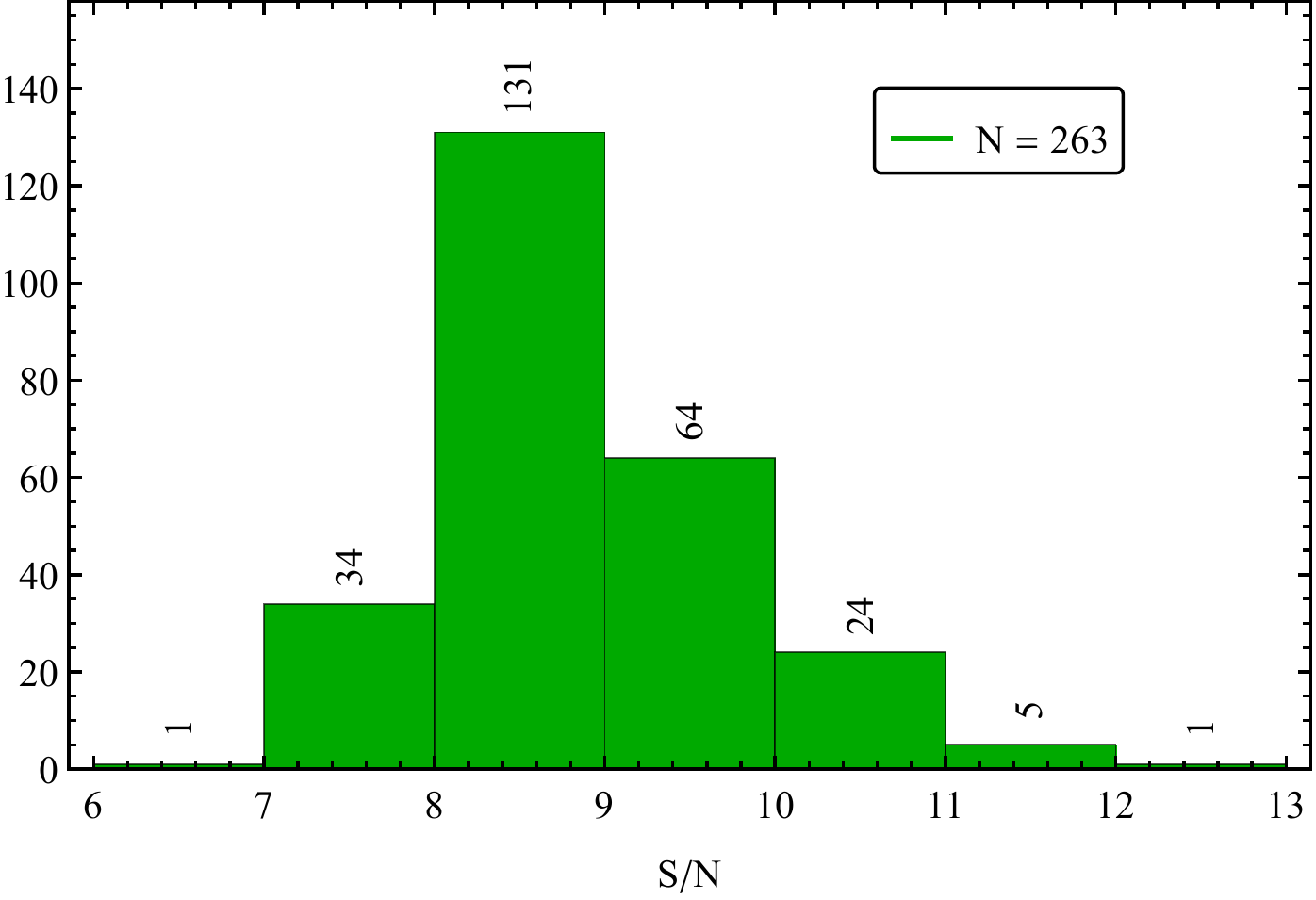}
\caption{\label{fig:EchoBackHist2} A histogram of the maximum SNR found within each background data set for the $\alpha = 2$ Boltzmann model. The measured signal SNR is again just under 12, leading to an estimated p-value of about 1 in 263. However, three anomalous points with SNR $> 12$, that we interpret as due to noise, were removed from the dataset.}
\end{figure}

Second, we compare this signal to 100 sets of noise backgrounds where it is known that no echoes exist, and find that the observed maximum SNR can be seen in 2/100 backgrounds, leading to a statistical significance of detection of about $2.3\sigma$ (Fig.\ref{fig:EchoBackHist}).
We repeat the same background comparison for the $\alpha = 2$ Boltzmann model, with 263 background searches and show the results in Fig.\ref{fig:EchoBackHist2}.

\begin{figure}
\centering
\includegraphics[width=0.47\textwidth]{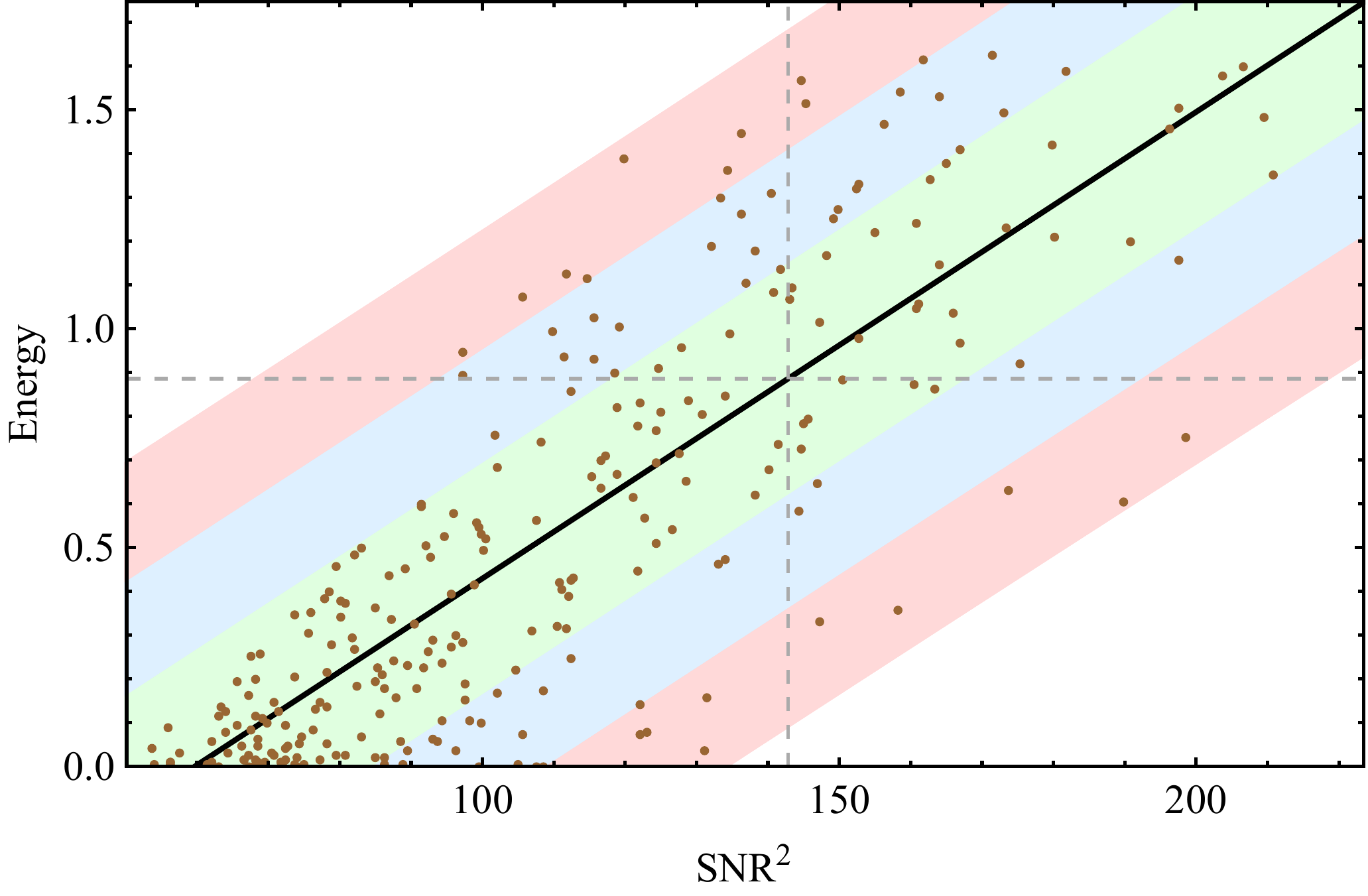}
\caption{\label{fig:InjCheckEvSNR} To calibrate the variation due to noise present in the linear model for energy as a function of maximum SNR, here we perform 263 injections in 263 sets of background data with a different energy injection for every background for $\alpha = 1$ Boltzmann echoes. The injections are ordered, starting at the first background with 0 energy and increasing in energy with each background up to an injection amplitude equal to $1.28\sigma$, where here $\sigma$ is calculated from the amplitudes of the Livingston data. The black line marks the linear model fit, with the vertical grey dashed line marking the value of the squared SNR seen in the echo data search and the horizontal line marking the corresponding measurement of the energy. The shaded regions mark the 1, 2, and 3$\sigma$ deviations away from the linear model. In this way, the echo data search and measured maximum SNR value can be seen as a measurement of echo energy at a value of $\sim 0.9^{+0.8}_{-0.8}$, where the errors mark the 99.7\% credibility region. Note: For the calculation of the linear model, three deviant points at high energy and high SNR were removed to improve the fit. Scaled by the GR event energy, this gives $E_{Echoes} / E_{GR} = 8.9 \pm 4.5\%$, where the uncertainty range represents the 90\% credible region. When repeating the same analysis for $\alpha = 2$ Boltzmann echoes, we find the same value for the energy ratio, but with $\pm 8.1\%$ uncertainty.}
\end{figure}

\begin{figure}
\centering
\includegraphics[width=0.47\textwidth]{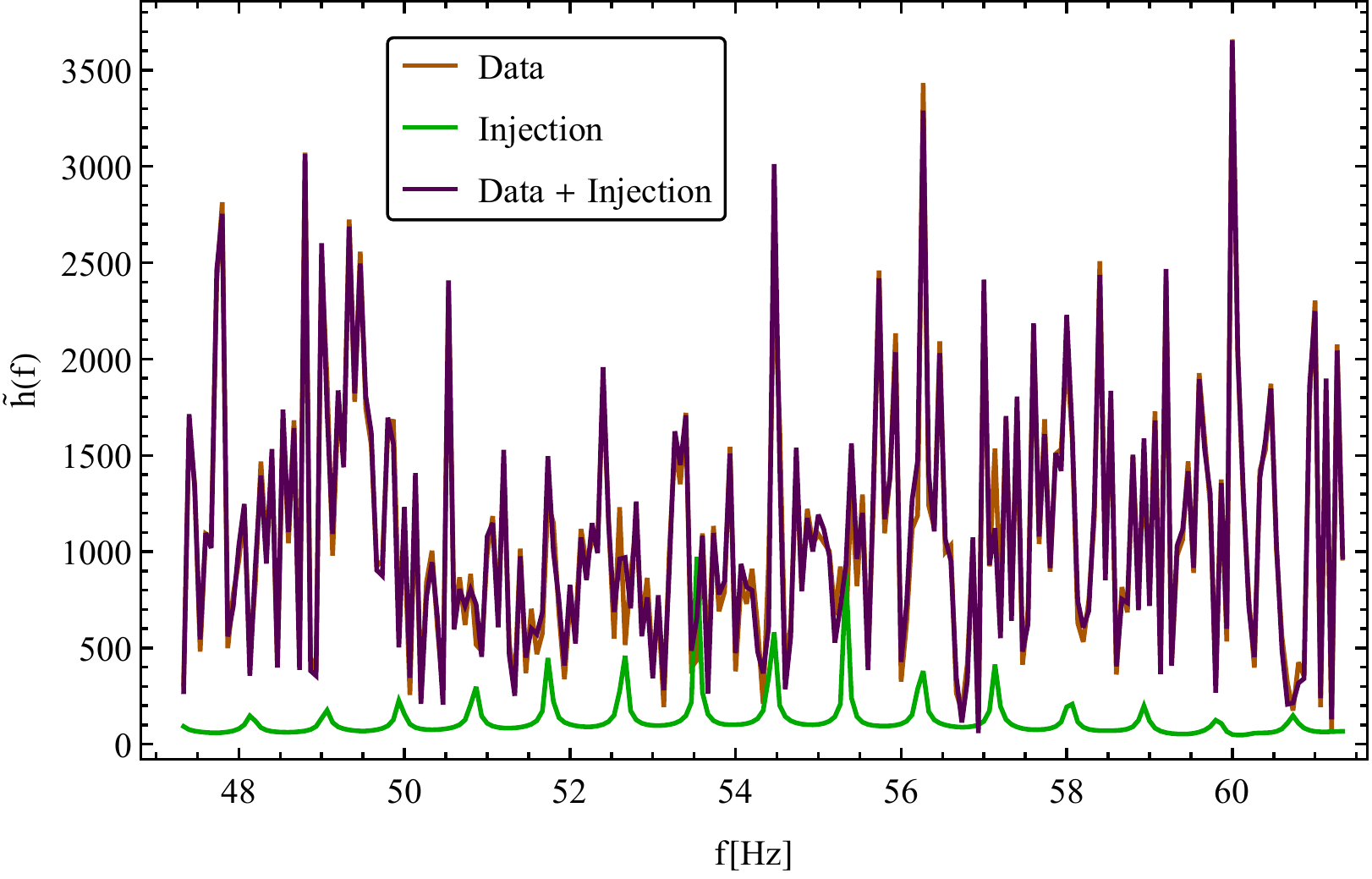}
\caption{\label{fig:MeasuredEnergyInj} From the calculated function for echo energy as a function of SNR, we can determine an approximate amplitude for the echoes required to obtain the measured SNR maximum of just under 12 in the data search. Shown here is the absolute value of the echo spectrum superimposed with the absolute value of the data before and after injection, for the Livingston detector. This plot assumes the most probable noise distribution where the maximum SNR prior to injection is 8.5.}
\end{figure}

Third, we characterize signal detection by injecting noise into 263 backgrounds with a different amplitude for each background. Since energy is the integral of $\omega^2 |f(\omega)|^2$, we relate the amplitude of these injections to the energy of the injected echoes. Because the SNR$^2$ tends to increase linearly with the square of the injection amplitude, we are then able to plot the energy of echoes as a function of SNR$^2$  (Fig.\ref{fig:InjCheckEvSNR}). 
The variation of the noise naturally present in the 263 unique background data sets provides error margins in the linear fit. Through this process, we find the most likely SNR for a given data set, and with the assumption that the signal data set features typical noise, we measure the energy of the echoes as the value of the linear regression at the measured SNR (see Fig.\ref{fig:MeasuredEnergyInj}).

\begin{figure}
\centering
\includegraphics[width=0.47\textwidth]{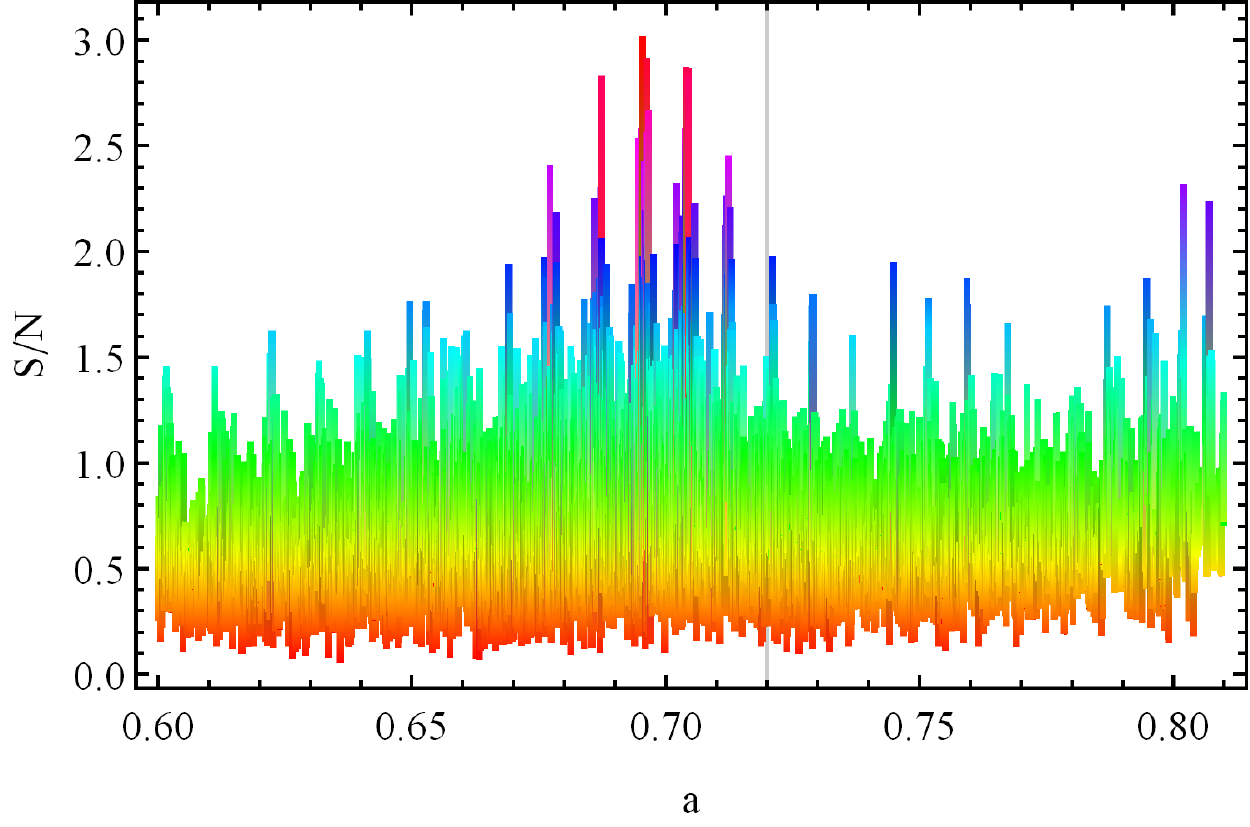}
\caption{\label{fig:LongTimeBoltzEchoSearch} Testing the robustness of the signal found in the standard Boltzmann echo data search, we also perform the same search but using the long time segment of $120s$, showing that a distinct signal is maintained.}
\end{figure}

\begin{figure}
\centering
\includegraphics[width=0.47\textwidth]{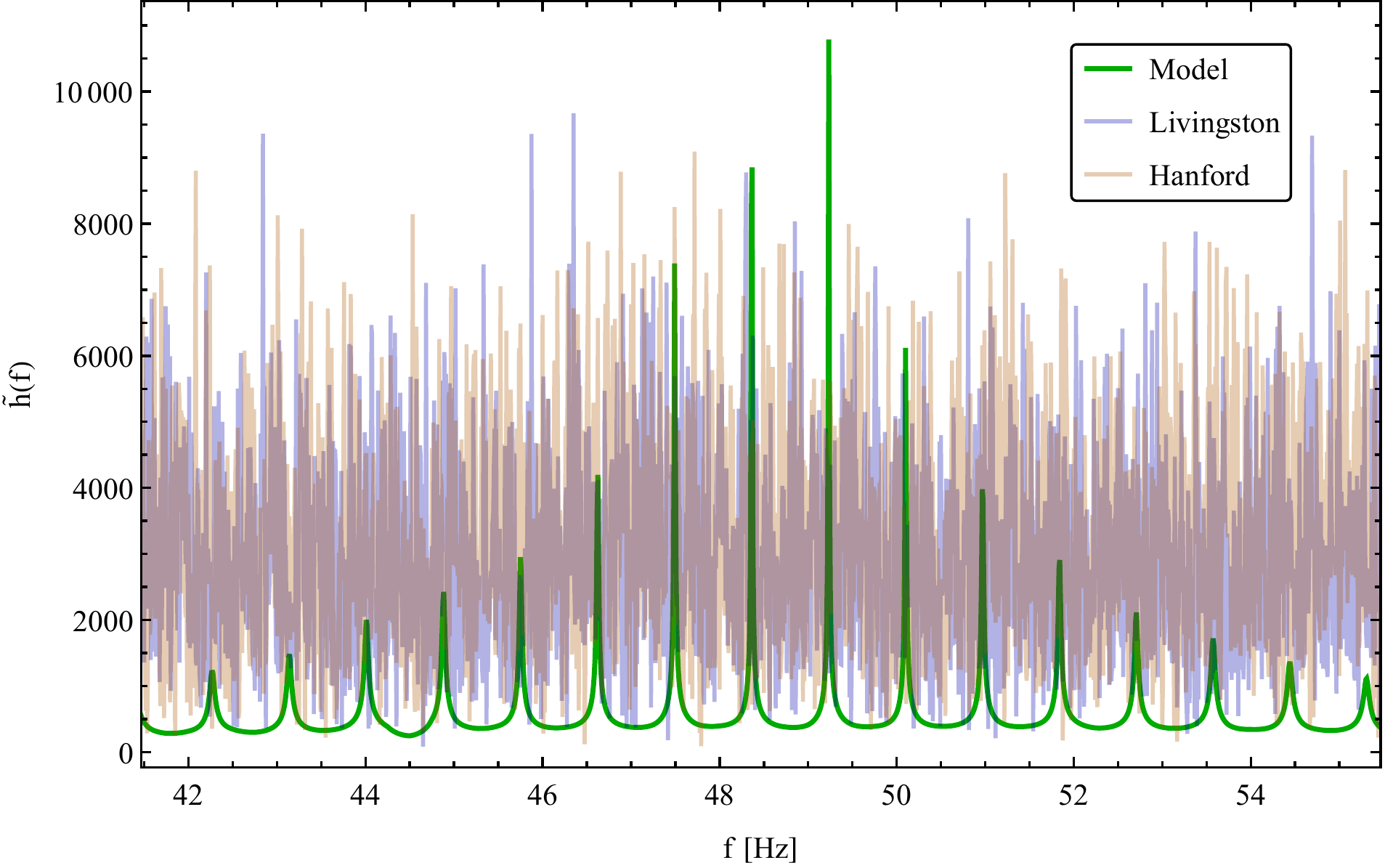}
\caption{\label{fig:LongTimeBoltzA2} The Boltzmann echoes primarily discussed in this paper assume standard reflectivity with $\alpha = 1$ \cite{S_Wang_40}, however this parameter has some flexibility, and here we test the value of $\alpha = 2$, which reduces the exponential suppression of the reflectivity by dividing the exponent by 2, and is the limiting value that preserves stability. We also maintain the long-duration time segment of $120s$. Compared to the standard shorter-time Boltzmann search, this result features a wider range of sharper resonances.}
\end{figure}

\begin{figure}
\centering
\includegraphics[width=0.47\textwidth]{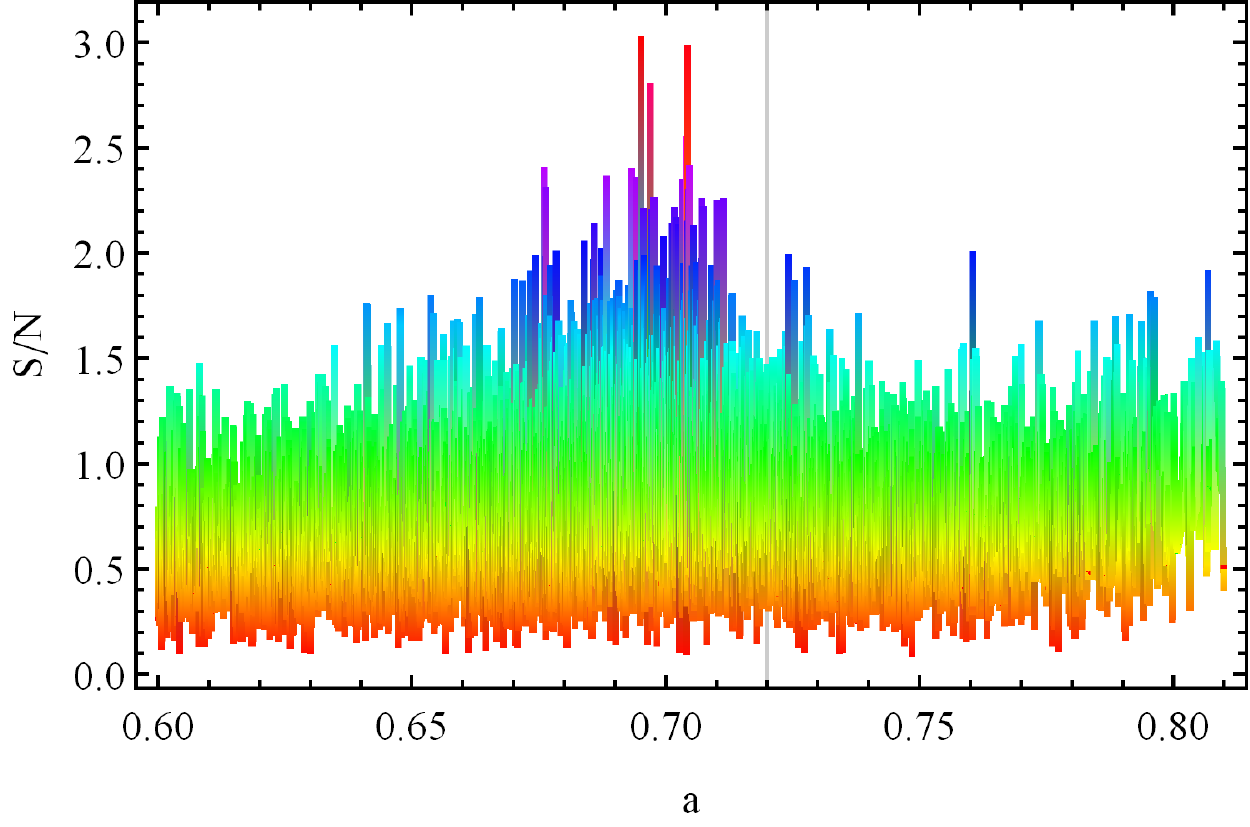}
\caption{\label{fig:LongTimeBoltzEchoSearchA2} Lastly, we repeat the original data search but with the $120s$ segment and the Botlzmann model constructed with $\alpha = 2$. Once again a signal is seen, and the noise is further suppressed away from the peak.}
\end{figure}

Finally, we make a couple moves away from our main analysis to test the robustness of the signal against variations in the model and search method. We find that a signal is preserved through changing the data length from $T = 15s$ to $T = 120s$ and the Boltzmann parameter from $\alpha = 1$ to $\alpha = 2$ (see Fig.\ref{fig:LongTimeBoltzEchoSearch}, Fig.\ref{fig:LongTimeBoltzA2}, and Fig.\ref{fig:LongTimeBoltzEchoSearchA2}). 
These are not fine-tuned parameters, but rather motivated by the long-time searches found in the literature \cite{S_Conklin_73, S_Holdom_90}.

\section{Conclusions}

We have presented a physically motivated surrogate model for Boltzmann echoes and deployed this in a data search. The Boltzmann model is suggested by potential quantum effects taking over within a Planck distance from the horizon radius, and we include this as our exotic echo-producing boundary condition. We represented the Boltzmann model as a sum over quasinormal modes parametrized by the location, amplitude, width, and phase of the resonances. We carefully examined each of these parameters and found numerically efficient and accurate representations for them through which we developed a computationally cheap surrogate model for Boltzmann echoes. For the separation function in particular, which is the function that determines the spacing of the characteristic echo resonances, we quantified a quartic order departure from the often-assumed constant $1 / t_d$ spacing with magnitude variation up to the order of several percent. To source our echoes, we use a numerical relativity surrogate model for the inspiral-merger-ringdown of a binary black hole system. This allows us to reproduce the observed waveform for GW190521, an event of particular interest for echo research due to its loud ringdown, as well as to determine the echoes that would be generated from this waveform. Finally, with the surrogate model in hand, we prepare the data surrounding GW190521 for analysis and perform a data search where we calculate the signal-to-noise ratio (SNR) over a range of Monto Carlo sets of mass and spin values distributed according to the most probable range of background values as measured by the LIGO and Virgo observatories. We find an SNR peak with moderate significance, and interpret this as a measurement of echo energy by relating the observed signal to simulated signals coming from injections into backgrounds. With this new tool at hand, we anticipate the future work of applying this physically motivated surrogate model, along with variations accounting for uncertainties in the physics, to additional gravitational events to further quantify the significance of echoes, and what they may imply for the quantum nature of black holes.

\begin{acknowledgements}

We thank Jahed Abedi for providing the Monte Carlo parameter chains used in the data search, Luis Longo for the inspiralling particle source function used in the original analysis, and Naritaka Oshita for helpful suggestions regarding the width parameter. We also thank all the members of our weekly group meetings for their supportive discussion and continual patience through our many conversations. We also thank Bob Holdom for invaluable discussion and advice. NA is supported by the University of Waterloo, Natural Sciences and Engineering Research Council of Canada (NSERC) and the Perimeter Institute for Theoretical Physics.
Research at Perimeter Institute is supported in part by the Government of Canada through
the Department of Innovation, Science and Economic Development Canada and by the
Province of Ontario through the Ministry of Colleges and Universities.  This research has made use of data, software and/or web tools obtained from the GW Open Science Center (https://www.gw-openscience.org), a service of LIGO Laboratory, the LIGO Scientific Collaboration and the Virgo Collaboration. LIGO is funded by the U.S. National Science Foundation. Virgo is funded by the French Centre National de Recherche Scientifique (CNRS), the Italian Instituto Nazionale della Fisica Nucleare (INFN) and the Dutch Nikhef, with contributions by Polish and Hungarian institutes.
\end{acknowledgements}

\appendix 
\section{Model Parameters and Calibration}\label{sec:Appendix}

\subsection{Surrogate Coefficients}

Here we include some explicit valuations of the surrogate model parameters, with units suppressed for simplicity. The surrogate model can be constructed for any set of background parameters studied in this paper, and here we choose the representative values of $a = 0.7$ (implying $k = \omega - 0.408$) and $x_0 = -450$ (implying $t_d = 900$). Other values may be calculated by the same methods.
\begin{equation}
\begin{split}
\Delta l_r = & \frac{1}{t_d}\left(6.33 - 117\frac{k}{t_d} - 3.36*10^6\frac{k^2}{t_d^2} \right. \\
& \left. - 2.62*10^{10}\frac{k^3}{t_d^3} - 7.24*10^{13} \frac{k^4}{t_d^4}\right) .
\end{split}
\end{equation}
\begin{equation}
\begin{split}
a_r = &\frac{450}{t_d}(2.68*10^{-2}R(|k|) - 5.59*10^{-5} - 2.30*10^{-2}k \\
& - 4.03*10^{-2}k^2 + 0.510k^3 + 1.46k^4) .
\end{split}
\end{equation}
\begin{equation}
w_r = \frac{1}{t_d}(1.52*10^{-2} + 6.64|k| + 10.2k^2 + 30.4k^3) .
\end{equation}

\subsection{Conversions and $R_{\textrm{BH}}$, $T_{\textrm{BH}}$}

\begin{figure}
\centering
\includegraphics[width=0.47\textwidth]{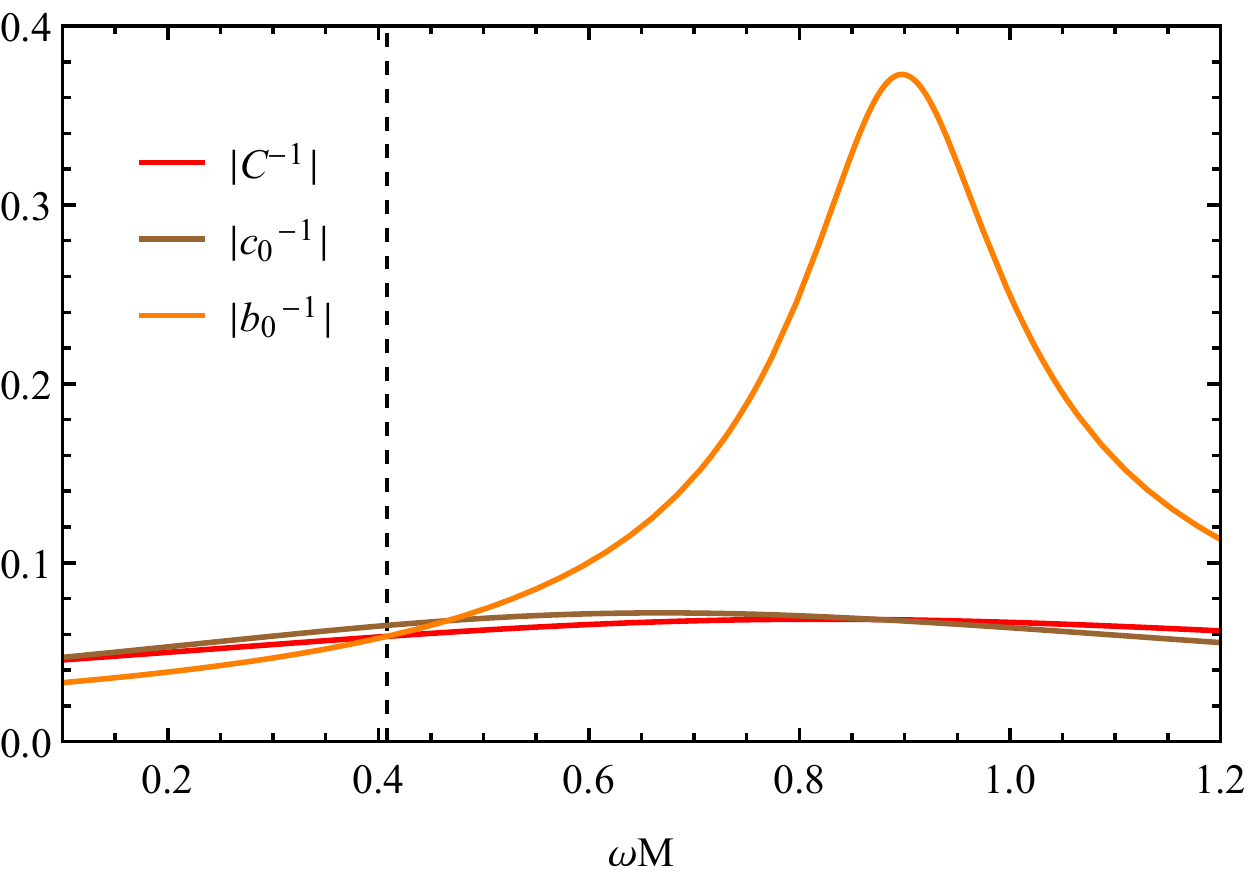}
\caption{\label{fig:ConversionFactors} Conversion factors present in the transformations between the strain, Teukolsky, and Sasaki-Nakamura bases. Neglecting normalization factors, the outgoing Teukolsky amplitude at infinity, $Z$, is related to the strain $h$ by $Z \sim \omega^2 h e^{-i\omega x_{obs}}$, where $x_{obs}$ is the observation point. To get the observed strain from the outgoing Sasaki-Nakamura amplitude at infinity, $X$, the conversion is $h \sim -\frac{4}{c_0} X e^{i\omega x_{obs}}$. Other amplitude and energy conversion factors and their uses are defined in \cite{S_Conklin_73}.}
\end{figure}

Several conversion factors required to transform between the homogenous asymptotic spectral amplitudes for the Teukolsky and the Sasaki-Nakamura formalisms are well known as formulas, and in Fig.\ref{fig:ConversionFactors} we plot them for convenience.

\begin{figure}
\centering
\includegraphics[width=0.47\textwidth]{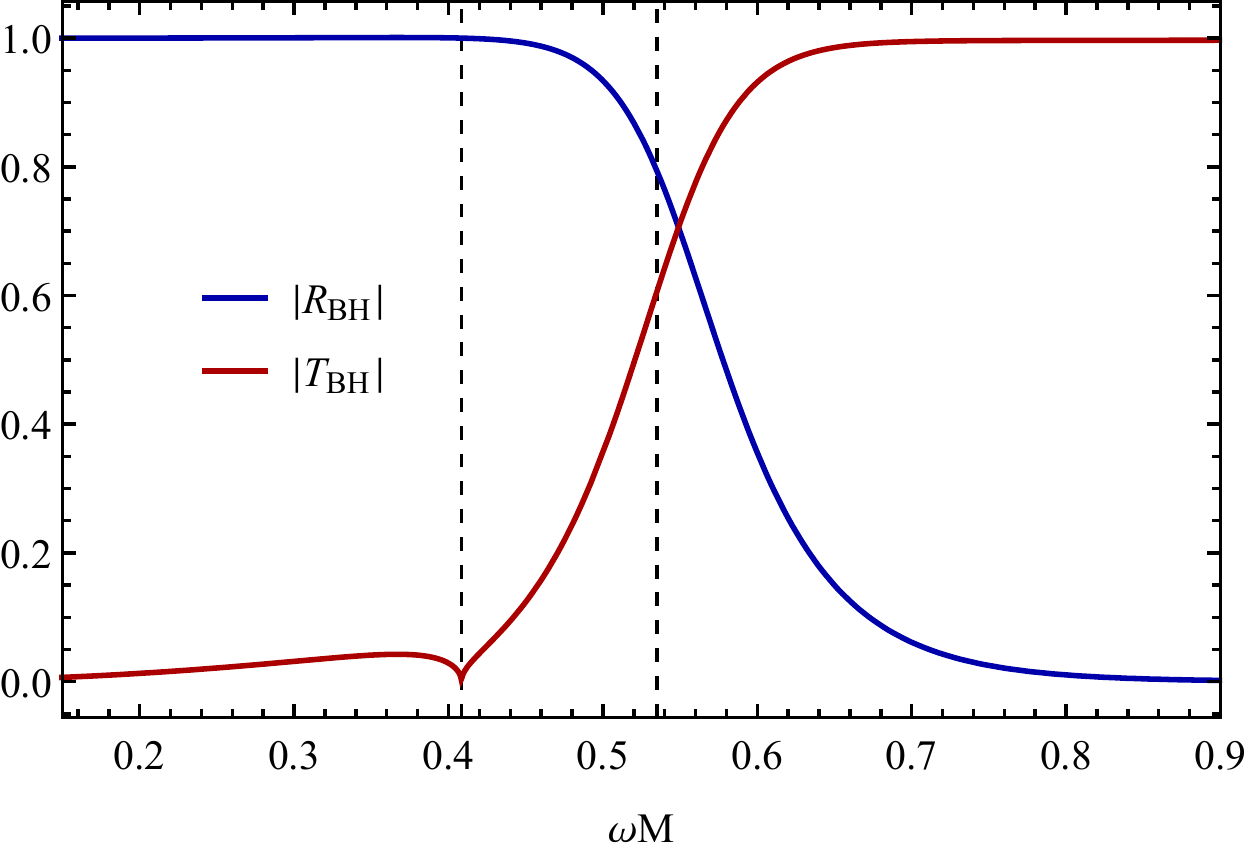}
\caption{\label{fig:RBHTBH} In red and blue, the absolute value of the transmission and reflection coefficients $T_{\textrm{BH}}(\omega)$ and $R_{\textrm{BH}}(\omega)$, respectively. The black dashed lines on the left and the right are the superradiance and fundamental QNM frequencies, respectively. The transmission function goes to zero and the reflection function goes to unity at the superradiance frequency. }
\end{figure}

The transmission and reflection coefficients $T_\textrm{BH}$ and $R_\textrm{BH}$, respectively, are frequently occurring in echo studies. This is especially true since they are independent of the boundary condition and are always present in some way in the transfer function. We plot them here in Fig.\ref{fig:RBHTBH}.
A couple key points should be mentioned here. First, the sum in quadrature of the absolute values of these is approximately unity everywhere. Second, the transmission coefficient goes to zero at the superradiance frequency - hence the spike in the source function at this frequency that reproduces the GR event (see Fig.\ref{fig:Source}) - and remains very low below this frequency. With these features in mind, it is possible to gain some intuition behind the geometric optics approximation for Boltzmann echoes between the Boltzmann boundary condition and the angular momentum barrier as a high pass filter.

\end{document}